\documentclass[12pt,oneside,fleqn]{article}

\usepackage[justification=centering]{caption}
\usepackage{array}
\usepackage{booktabs}
\usepackage{amsmath}
\usepackage{amssymb}
\usepackage{amsfonts}
\usepackage{amstext}
\usepackage{amsbsy}
\usepackage{graphicx}
\usepackage{subfigure}
\usepackage{xcolor}
\usepackage{natbib}
\usepackage{pdflscape}
\usepackage{longtable}
\usepackage{arydshln}
\usepackage{multirow}

\newcommand{\beq}{\begin{equation}}
\newcommand{\eeq}{\end{equation}}

\newcommand{\bz}{\mathbf{z}}

\newcommand{\bg}{\mathbf{g}}
\newcommand{\bh}{\mathbf{h}}
\newcommand{\bp}{\mathbf{p}}
\newcommand{\bt}{\mathbf{t}}
\newcommand{\bu}{\mathbf{u}}
\newcommand{\bw}{\mathbf{w}}
\newcommand{\bv}{\mathbf{v}}
\newcommand{\bx}{\mathbf{x}}
\newcommand{\by}{\mathbf{y}}

\newcommand{\bX}{\mathbf{X}}
\newcommand{\bW}{\mathbf{W}}

\newcommand{\bB}{\mathbf{B}}

\newcommand{\bE}{\mathbf{E}}
\newcommand{\bF}{\mathbf{F}}
\newcommand{\bP}{\mathbf{P}}
\newcommand{\bQ}{\mathbf{Q}}
\newcommand{\bA}{\mathbf{A}}

\newcommand{\bR}{\mathbf{R}}
\newcommand{\bS}{\mathbf{S}}
\newcommand{\bT}{\mathbf{T}}
\newcommand{\bU}{\mathbf{U}}
\newcommand{\bV}{\mathbf{V}}
\newcommand{\bZ}{\mathbf{Z}}
\newcommand{\bD}{\mathbf{D}}
\newcommand{\bI}{\mathbf{I}}

\newcommand{\mytoprule}{\specialrule{0.1em}{0em}{0em}}
\newcommand{\mybottomrule}{\specialrule{0.1em}{0em}{0em}}

\begin{document}

\begin{center}
{\Large
Common and Distinct Components in Data Fusion
}
\end{center}

\noindent Age K. Smilde\footnote{Author for correspondence. e-mail a.k.smilde@uva.nl}$^{\dag }$,
Ingrid M{\aa}ge$^{2}$, Tormod Naes$^{2}$, Thomas Hankemeier$^{3}$, Mirjam A. Lips$^{4}$, Henk A.L. Kiers$^{5}$, Evrim Acar$^{6}$ and Rasmus Bro$^{6}$
\newline

\begin{itemize}
{\footnotesize
\item[$^{1}$]
Biosystems Data Analysis, Faculty of Sciences, University of Amsterdam, Science Park 904, P.O. Box 94215,
1090 GE Amsterdam, The Netherlands
\item[$^{2}$]
Nofima, {\AA}s, Norway 
\item[$^{3}$]
LACDR, Leiden University, Netherlands
\item[$^{4}$]
Leiden University Medical Center, Department of Endocrinology and Metabolism, Netherlands
\item[$^{5}$]
Heymans Institute, University of Groningen, Netherlands
\item[$^{6}$]
Department of Food Science, University of Copenhagen, Denmark
}
\end{itemize}

\begin{abstract}
In many areas of science multiple sets of data are collected pertaining to the same system. Examples are food products which are characterized by different sets of variables, bio-processes which are on-line sampled with different instruments, or biological systems of which different genomics measurements are obtained. Data fusion is concerned with analyzing such sets of data simultaneously to arrive at a global view of the system under study. One of the upcoming areas of data fusion is exploring whether the data sets have something in common or not. This gives insight into common and distinct variation in each data set, thereby facilitating understanding the relationships between the data sets. Unfortunately, research on methods to distinguish common and distinct components is fragmented, both in terminology as well as in methods: there is no common ground which hampers comparing methods and understanding their relative merits. This paper provides a unifying framework for this subfield of data fusion by using rigorous arguments from linear algebra. The most frequently used methods for distinguishing common and distinct components are explained in this framework and some practical examples are given of these methods in the areas of (medical) biology and food science.\\


\end{abstract}

\section{Introduction and Motivation}
\label{Introduction and Motivation}

\subsection{Data fusion}

\noindent Simultaneous analysis of several data blocks has been proposed a long time ago \cite{Kettenring1971,VandeGeer1984}, but today we can see a renewed interest fueled by the strongly increasing needs in many sciences. A number of different methods have been put forward \cite{Alter2003,Trygg2003,VanDeun2012,Lock2013,Acar2014} all of them with a common interest of either understanding relations better or obtaining better prediction results. The methodologies are known under different names in different disciplines, important examples being data fusion, data integration, multi-block analysis, multi-set analysis and multi-mode analysis (for definitions, see \cite{VanMechelen2010,Lahat2015}). Some of the methods are rather straightforward generalizations of standard methods for one or two data sets such as concatenated PCA and PLS regression \cite{Westerhuis1998}, while others are explicitly developed for handling multi-block data focusing on a number of concepts unique for such applications.  In the latter group one can find methods such as SO-PLS and PO-PLS \cite{Naes2013}, DISCO-SCA \cite{Schouteden2013}, O2PLS \cite{Trygg2003} and GSVD \cite{Golub1996}.\\

\noindent This paper will focus on one particular aspect that appears crucial in data fusion, namely the distinction between common and distinct information in the blocks. The main aim is to provide concrete definitions of the concepts and to discuss how these definitions relate to the most well known methods in the area. Main attention will be given to interchangeable data blocks sharing the row-mode which usually consists of samples or subjects; thus multi-block predictive methods such as SO-PLS, PO-PLS \cite{Naes2013} are not discussed. We will also restrict ourselves to direct analysis in contrast to indirect analysis such as analyzing covariance or correlation matrices. Focus will be on definitions based on column spaces: the spaces spanned by object scores on the variables but interpretation in terms of variable loadings (i.e. the row-space) will also be given some attention. Selected methods will be illustrated by real data sets. Situations with two blocks as well as situations with more than two blocks will be discussed. As an integral part of the discussion, we will also incorporate relative measures of fit of the different parts of the blocks.

\subsection{Motivating examples}

\subsubsection{Food Science}

In food product development we are typically interested in understanding how product formulations (ingredients etc.) of a set of product prototypes are related to the descriptive sensory properties of the product and also possibly to the consumer liking of the product. A typical situation might be that one is interested in substituting one of the ingredients by a cheaper one and is interested in seeing whether this change has any noticeable effect on the smell, the taste, the texture or all of them. Another typical situation is in new product development where the developer wants to understand how two important sensory modalities such as smell and taste are affected by the ingredients used. In both cases, it is crucial for further product optimization to know how this happens, for instance whether the smell and taste have a joint source of variability and/or what is influencing only one of them.

\subsubsection{Biology}

An important class of health problems is Diabetes Mellitus Type II (DM2). Consider measurements performed on a group of DM2 patients using a metabolomics platform (e.g. LC-MS), clinical measurements (such as insulin resistance, fasting glucose levels, blood pressure) and life-style variables. Then these measurements will have parts in common and have distinctive parts. The common part between the metabo-lomics and clinical measurements may reflect the relation between branched amino-acids and insulin resistance \cite{Lynch2014}; there may also be common parts between the life-style variables and the clinical measurements, such as exercise and blood pressure. Some of the metabolites, such as bile acids, may not be directly related to insulin resistance and life-style and will, hence, be distinct. Since all measurements pertain to the same system (DM2) it is worthwhile exploring and understanding the complete data set in a holistic way.

\subsubsection{General idea}

\begin{figure}[h!]
 \centerline{\includegraphics[width=12cm]{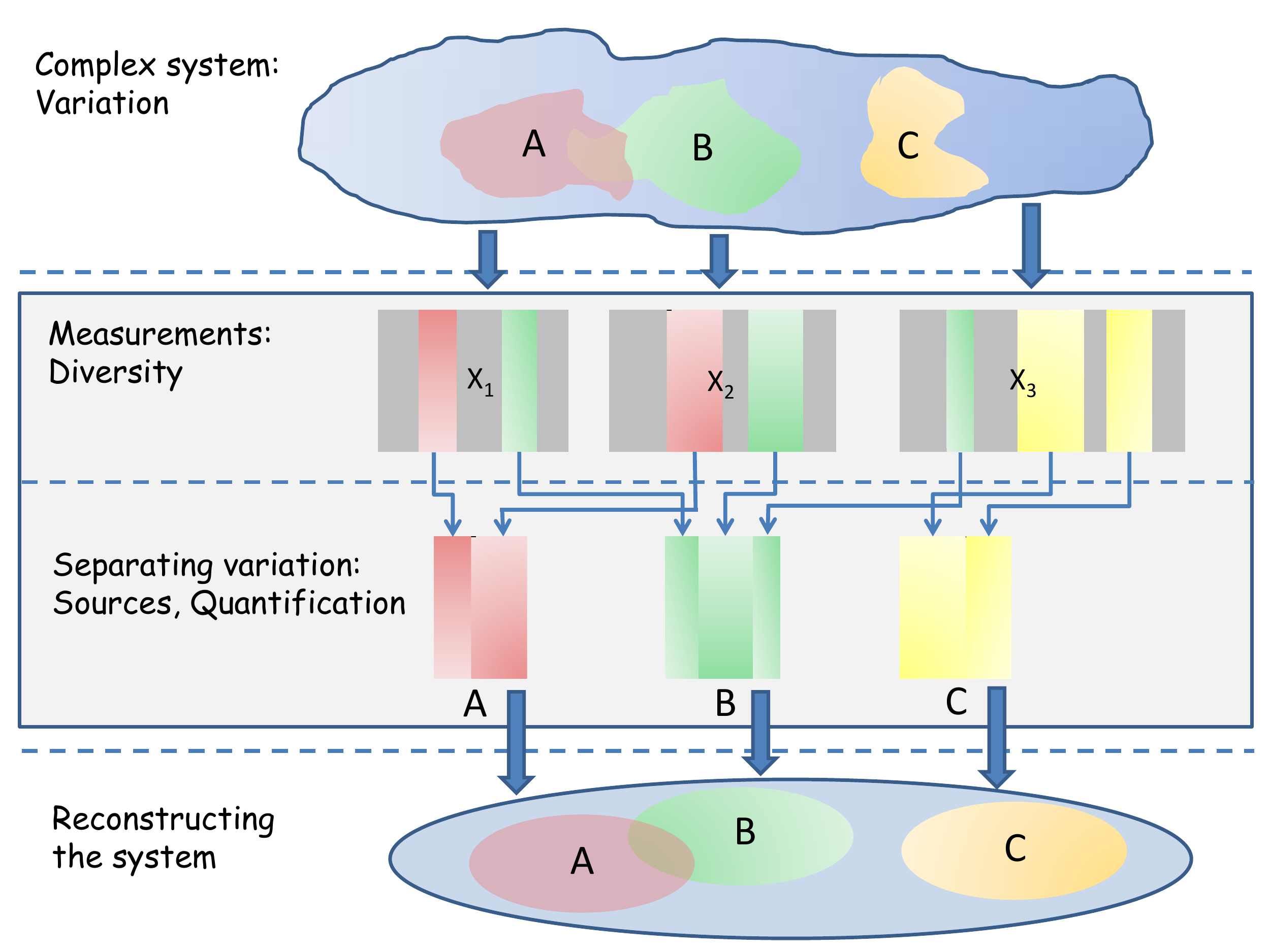}}
 \caption{\footnotesize Measurements are performed on a complex system probing parts A, B and C of that system. The resulting data blocks $\bX_1$, $\bX_2$ and $\bX_3$ contain mixed variation which has to be separated in sources  (red/green is common; yellow is distinct; grey is irrelevant variation and noise). These quantified sources are then used to reconstruct the system. This paper concerns mainly the box within the blue lines.}
 \label{Fig_Cartoon}
\end{figure}

The above two examples show common features which are summarized in Figure \ref{Fig_Cartoon}. Knowledge is required of a complex system (first and upper layer; e.g. DM2). Measurements are performed on this system resulting in three blocks of data $\bX_1$, $\bX_2$ and $\bX_3$ (second layer; e.g. metabolomics, clinical and life-style measurements in the DM2 example; smell, taste and consumer liking in the food science example). These measurements are preferably collected in such a way that \emph{diversity} is increased (\cite{Sidiropoulos2000,Lahat2015}. Although diverse and information-rich data is obtained, the problem is that the data blocks contain partly overlapping contributions of parts A, B and C of the system (e.g. A is insulin-glucose-amino-acid metabolism and B reflects cardiocasvular complications in the DM2 example; sweetness (A) in the case of taste and smell in the food science example) and also irrelevant variation and noise. The idea behind finding common and distinct variation in the three data blocks is to separate and quantify the different sources of variation which are spread across all data blocks (third layer). Interpreting the different sources of variation will then lead to a reconstruction of the system (fourth and bottom layer; e.g. the etiology of DM2). In our paper, we will mainly describe moving from the second to the third layer (the boxed part), and will only touch upon moving from the third to the fourth layer. In Section \ref{Examples} we will present some real-life examples which were already introduced above.

\section{General mathematical framework}

\noindent For the definition of the basic concepts we will start with two data matrices or blocks $\bX_1$ of size ($I \times J_1$) and $\bX_2$ of size ($I \times J_2$) and afterwards discuss how these concepts can be extended to three or several blocks of data. It is assumed that the two matrices share the first mode (the \textit{I}-mode, \cite{Tauler1995}) usually representing samples or objects and the data have been column-centered throughout. Note that the two data sets may have different number of columns, usually representing variables, which means that they may (and often will) contain different types of measurements.\\

\noindent This section will be devoted to precise definitions of common and distinct components for the blocks in the data set. All these definitions are inspired by and related to previous definitions, but the main aim here is to make the definitions precise and unambiguous and therefore better suited for comparing methodologies.  The definitions will be made in terms of subspaces, but later on we will expand to discuss the same concepts in terms of components which are basis vectors, chosen in one way or another, for the subspaces.\\

\noindent The mathematical framework represents the idealized situation of noiseless data. In practice, of course, this never happens. Hence, in later Sections we are also going to discuss which kind of compromises and choices have to be made in real-life situations. In that context, we also discuss several existing methods for finding common and distinct subspaces as used in the psychometrics, bioinformatics, chemometrics, computer science, data analysis and statistics literature.

\subsection{Description of the framework}
\label{Description of the framework}

\subsubsection{The two-block case}
\label{The two-block case}

\begin{figure}[h!]
 \centerline{\includegraphics[width=10cm]{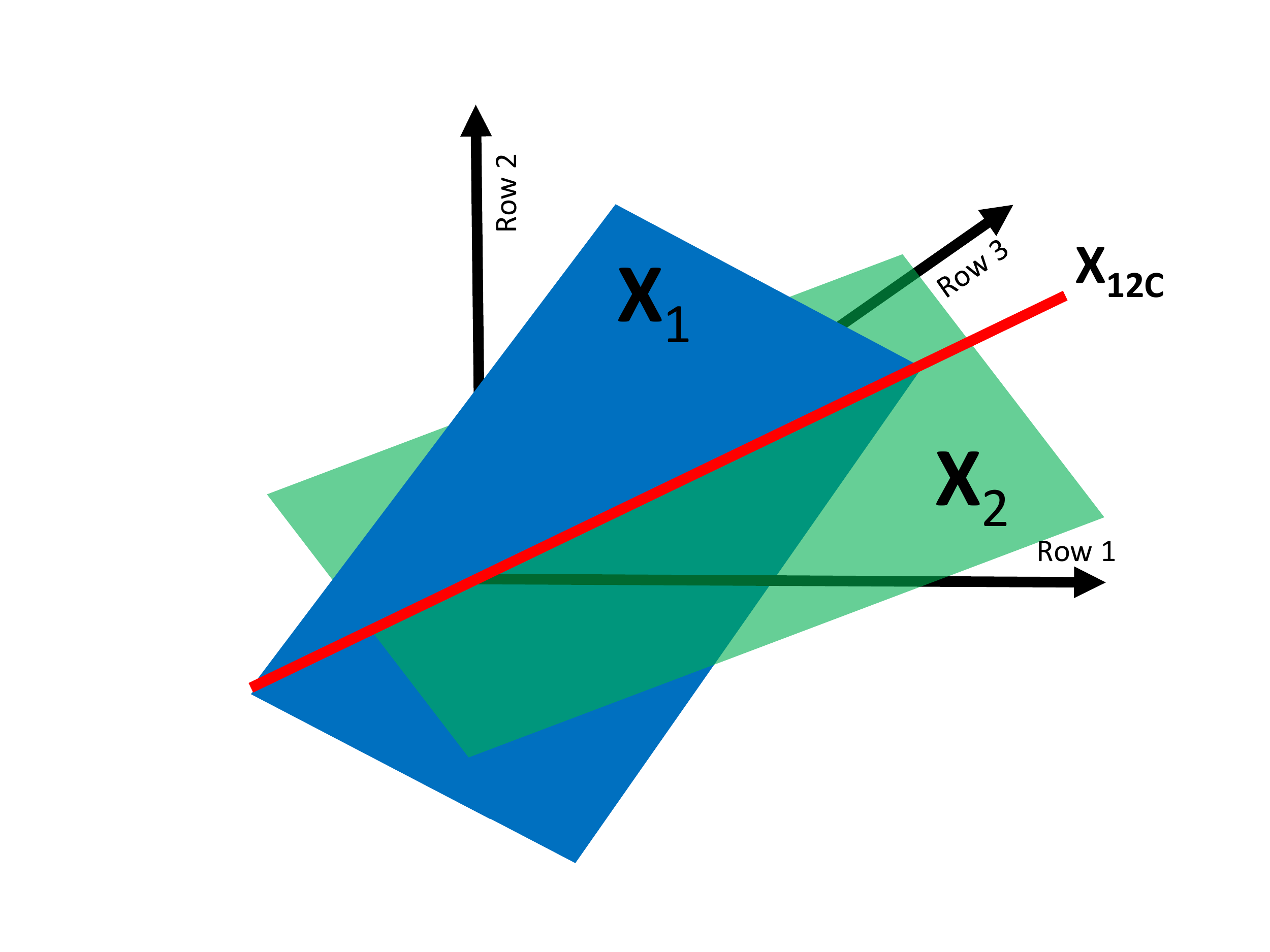}}
 \caption{\footnotesize The \textit{I}-dimensional space having $R(\bX_1)$ (blue) and $R(\bX_2)$ (green) as subspaces. Only three axis of this \textit{I}-dimensional space are drawn. The red line $\bX_{12C}$ represents the common subspace. For the sake of illustration the dimensions of both column-spaces are equal (two). This is not necessarily always the case.}
 \label{Fig_Common}
\end{figure}

\noindent The two spaces spanned by the columns of $\bX_1$ and $\bX_2$ ($R(\bX_1)$ and $R(\bX_2)$) are located in the same \textit{I}-dimensional column-space $R^I$, see Figure \ref{Fig_Common} for an illustration in three dimensional space. Each variable is a vector in this coordinate system indicating the level of that variable for each sample (row). These variables are not explicitly shown in this figure but will lie within the space indicated by the blue and green column-spaces.\\

\noindent If the two column-spaces intersect non-trivially (the zero is always shared), then the intersection space is called the common space. In Figure \ref{Fig_Common} there is only one common direction (i.e. the common space is one-dimensional), but there can be more or none. The common subspace will be called $R(\bX_{12C})$ where the subscript C stands for 'Common'. Note that $R(\bX_{12C})\subseteq R(\bX_1)$ and $R(\bX_{12C})\subseteq R(\bX_2)$. The common part of the two blocks will in most cases not span the whole of $R(\bX_1)$ and $R(\bX_2)$. Some definitions regarding the rest of these spaces are therefore needed. As will be discussed later, it is useful to distinguish between different ways of representing these subspaces, depending on choices regarding orthogonality. In all cases, these subspaces representing the rest after identification of the common part will be called "distinct" subspaces. The requirement is that the space spanned by the columns in a block $\bX_k (k=1,2)$ is a direct sum of the common space and the distinct space within that block. Hence, these two parts within a block are linearly independent (two subspaces are linearly independent if no vector in one subspace can be written as a linear combination of the vectors of the other and vice versa).\\

\begin{figure}[h!]
 \centerline{\includegraphics*[width=14cm]{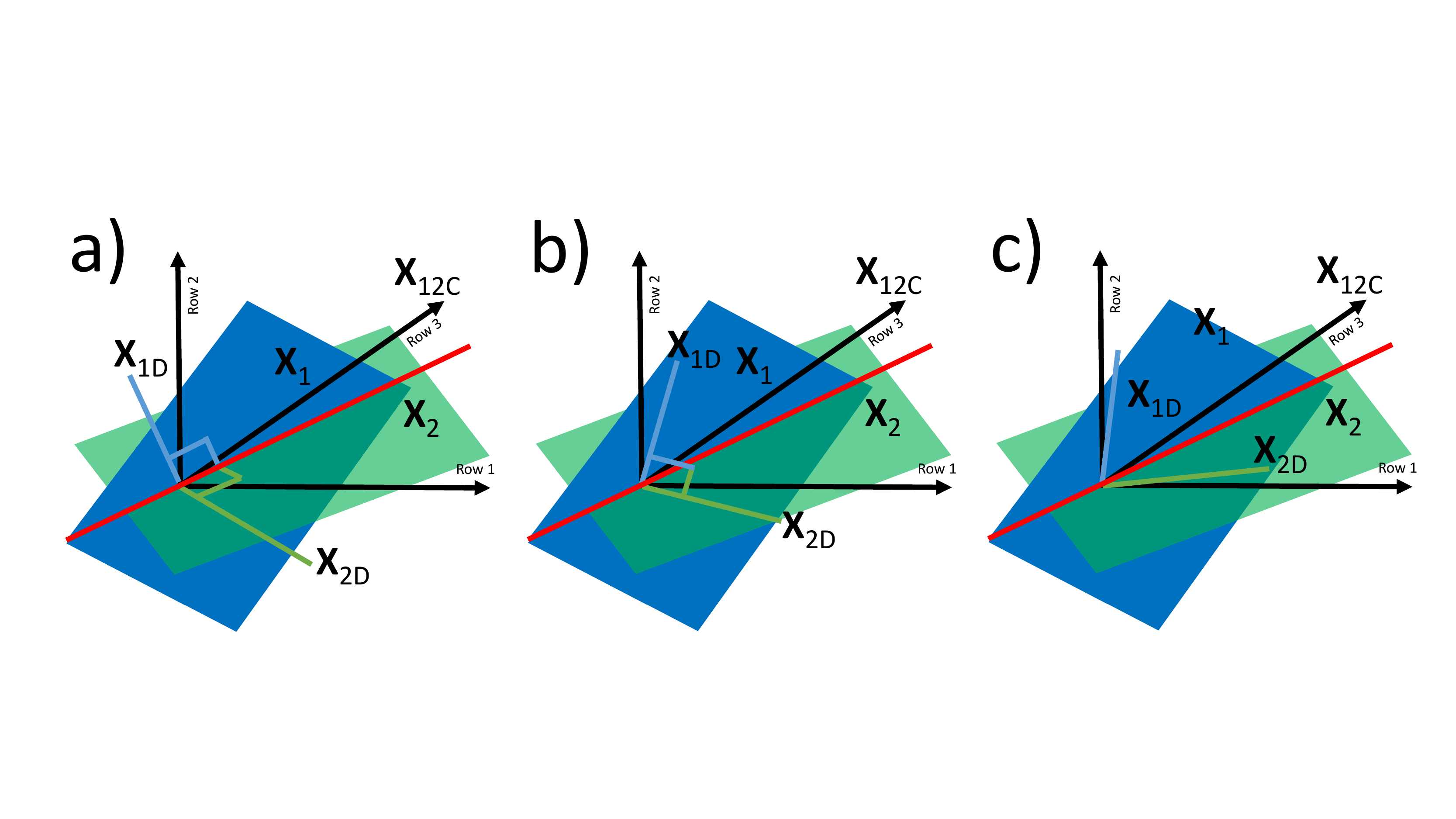}}
 \caption{\footnotesize See also legend Figure \ref{Fig_Common}. The distinct subspaces (1-dimensional in this case) are spanned by $\bX_{1D}$ and $\bX_{2D}$ for $R(\bX_1)$ and $R(\bX_2)$, respectively; a) both distinct subspaces are chosen orthogonal to the common subspace,  b) both distinct subspaces are chosen mutually orthogonal, c) no orthogonality.}
 \label{Fig_Distinct}
\end{figure}

\noindent These subspaces are called $R(\bX_{1D})$ and $R(\bX_{2D})$ where the subscript D stands for 'Distinct'. The choice whether or not to choose orthogonality depends on the application. In Figure \ref{Fig_Distinct} three possibilities are shown, namely making the distinct subspaces orthogonal to the common subspace or making the distinct subspaces orthogonal to each other or imposing no orthogonality at all. In general, it is not possible to combine the orthogonalities of Figure \ref{Fig_Distinct}a and b.\\

\noindent What we have accomplished now is decomposing $R(\bX_1)$ and $R(\bX_2)$ into direct sums of spaces:
\begin{eqnarray}\label{eDirectSum1}
  R(\bX_1) &=& R(\bX_{12C}) \oplus R(\bX_{1D}) \\ \nonumber
  R(\bX_2) &=& R(\bX_{12C}) \oplus R(\bX_{2D})
\end{eqnarray}
because $R(\bX_{12C}) \cap R(\bX_{1D})=\{0\}$ and $R(\bX_{12C}) \cap R(\bX_{2D})=\{0\}$ \cite{Schott1997}. Hence, it also holds that
\begin{eqnarray}\label{eDim1}
    dim R(\bX_1) &=& dim R(\bX_{12C})+ dim R(\bX_{1D}) \\ \nonumber
    dim R(\bX_2) &=& dim R(\bX_{12C})+ dim R(\bX_{2D})
\end{eqnarray}
If the distinct-orthogonal-to-common option is chosen (see Figure \ref{Fig_Distinct}a), then additionally it holds that $R(\bX_{12C}) \bot R(\bX_{1D})$ and $R(\bX_{12C}) \bot R(\bX_{2D})$. Note that for this case, given the common space, the decomposition is unique because then $R(\bX_{1D})$ is the orthogonal complement of $R(\bX_{12C})$ within $R(\bX_1)$ and likewise for $R(\bX_{2D})$ (but not necessarily the basis within the subspaces if these have dimension higher than one). In the non-orthogonal case, the distinct part can be defined by any set of linearly independent vectors that are in the original spaces, but not in the common space. For a thorough description of direct sums of spaces, see \cite{Yanai2011}.\\

\noindent We can take it one step further by also decomposing both the distinct subspaces $R(\bX_{1D})$ and $R(\bX_{2D})$ in two parts:
\begin{eqnarray}\label{eDistinct2}
    R(\bX_{1D}) &=& R(\bX_{1DO}) \oplus R(\bX_{1DNO}) \\ \nonumber
    R(\bX_{2D}) &=& R(\bX_{2DO}) \oplus R(\bX_{2DNO})
\end{eqnarray}
where $R(\bX_{1DO})$ is the "distinct-orthogonal (DO)" part and the other part will be called "distinct-non-orthogonal" (DNO); where $R(\bX_{1DO}) \bot R(\bX_{2DO})$ and $R(\bX_{1DNO})$ is the remaining part of $R(\bX_{1D})$ after removing $R(\bX_{1DO})$ and likewise for $R(\bX_{2DNO})$. Again, by the definition of direct sum we have $R(\bX_{1DO}) \cap R(\bX_{1DNO})=\{0\}$ and $R(\bX_{2DO}) \cap R(\bX_{2DNO})=\{0\}$. The argument for the split of Eqn. \ref{eDistinct2} is that one may be interested in looking at the parts of the blocks that have no correlation with (parts of) each other at all. Note that such an additional split can only be performed when the dimensions of the subspaces allow so, e.g., in Figure \ref{Fig_Distinct} both distinct subspaces have only dimension one and thus cannot be decomposed further. Depending on the dimensions of the distinct subspaces and their relative positioning in space, different possibilities can be distinguished. A choice has to be made by the user and is application dependent. A summary of alternatives is presented in the Appendix (Section \ref{Possibilities of orthogonal decompositions}) but one example is the following. If $\bX_1$ and $\bX_2$ contain measurements of two instruments, then choosing $R(\bX_{2DO})$ orthogonal to the whole of $R(\bX_1)$ can be interpreted as the unique contribution of instrument 2 relative to instrument 1 or, stated differently, what is the gain by adding instrument 2?\\

\noindent Summarizing, we arrive at the following direct sum decompositions of the column-spaces of $\bX_1$ and $\bX_2$:
\begin{eqnarray}\label{eDirectSum2}
    R(\bX_1) &=& R(\bX_{12C}) \oplus R(\bX_{1D}) = R(\bX_{12C}) \oplus R(\bX_{1DO}) \oplus R(\bX_{1DNO}) \\ \nonumber
    R(\bX_2) &=& R(\bX_{12C}) \oplus R(\bX_{2D}) = R(\bX_{12C}) \oplus R(\bX_{2DO}) \oplus R(\bX_{2DNO})
\end{eqnarray}
representing our general definition of the basic concepts of common (C), distinct (D), distinct-orthogonal (DO) and distinct-non-orthogonal (DNO) subspaces. This decomposition is unique meaning that when the decomposition of Eqn. \ref{eDirectSum2} is chosen, then every vector in $R(\bX_1)$ can be written uniquely as a sum of three vectors in the three different subspaces $R(\bX_{12C})$, $R(\bX_{1DO})$ and $R(\bX_{1DNO})$ and likewise for $R(\bX_2)$, if the dimensions allow so.\\

\noindent The decomposition of Eq. \ref{eDirectSum2} gives also a break-down of the dimensions of the separate subspaces:
\begin{gather}\label{eDim2}
    dim R(\bX_1) = dim R(\bX_{12C})+ dim R(\bX_{1D})\\ \nonumber
    = dim R(\bX_{12C})+ dim R(\bX_{1DO}) + dim R(\bX_{1DNO})\\ \nonumber
    dim R(\bX_2) = dim R(\bX_{12C})+ dim R(\bX_{2D})\\ \nonumber
    = dim R(\bX_{12C})+ dim R(\bX_{2DO}) + dim R(\bX_{2DNO}).
\end{gather}

\subsubsection{Generalizations to three blocks}

The generalization to three blocks of data goes as follows. Consider the sets $\bX_1 (I \times J_1)$, $\bX_2 (I \times J_2)$ and $\bX_3 (I \times J_3)$. We can define again a part which is in common between all three column-spaces, $R(\bX_{123C})$ with obvious notation. Next, we can define a part in common between $R(\bX_1)$ and $R(\bX_2)$ which is not intersecting with $R(\bX_3)$, $R(\bX_{12C})$, and likewise we can define $R(\bX_{13C})$ and $R(\bX_{23C})$. The complete part of $R(\bX_1)$ which is shared with the other blocks can then be written as $R(\bX_{123C}) \oplus R(\bX_{12C}) \oplus R(\bX_{13C})$ with the properties that $R(\bX_{123C}) \cap R(\bX_{12C})= \{0\}$, $R(\bX_{123C}) \cap R(\bX_{13C})=\{0\}$ and $R(\bX_{12C}) \cap R(\bX_{13C})=\{0\}$.\\

\begin{figure}[h!]
 \centerline{\includegraphics*[width=14cm]{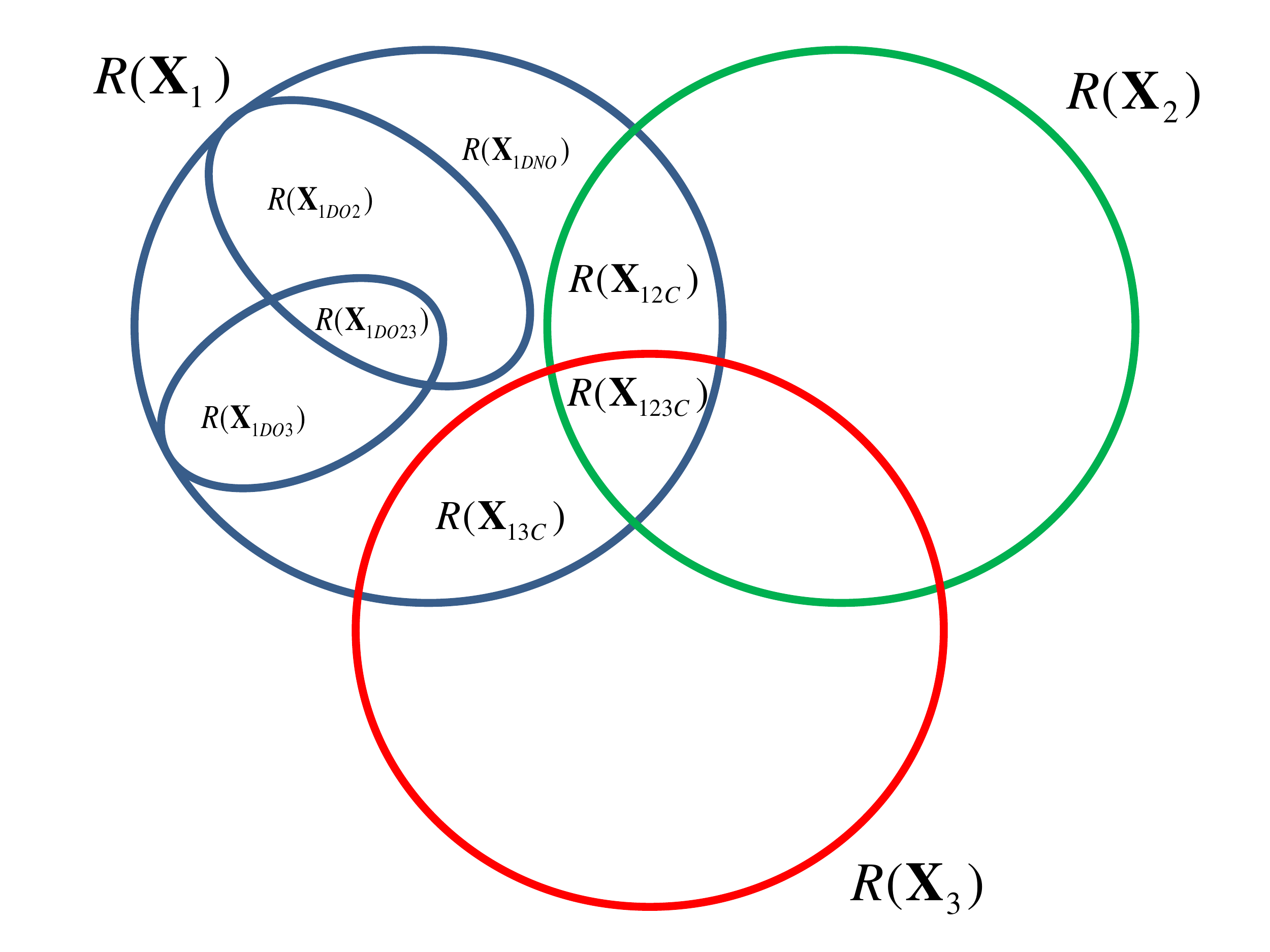}}
 \caption{\footnotesize The decomposition of $R(\bX_1)$ in the three block situation.}
 \label{Fig_ThreeBlocks}
\end{figure}

\noindent The distinct part of $R(\bX_1)$ can again be defined as the part of $R(\bX_1)$ linearly independent of $R(\bX_{123C}) \oplus R(\bX_{12C}) \oplus R(\bX_{13C})$. This leads to the following decomposition:
\begin{equation}\label{eDirectSum3}
    R(\bX_1)=R(\bX_{123C}) \oplus R(\bX_{12C}) \oplus R(\bX_{13C}) \oplus R(\bX_{1D})
\end{equation}
and the distinct part $R(\bX_{1D})$ can again be broken down in several parts. The first part may be chosen to be the subspace of $R(\bX_{1D})$ orthogonal to $R(\bX_2) \cup R(\bX_3)$ with obvious notation $R(\bX_{1DO23})$. Then there is a part orthogonal to only $R(\bX_2)$, $R(\bX_{1DO2})$, and a part only orthogonal to only $R(\bX_3)$, $R(\bX_{1DO3})$, where again $R(\bX_{1DO23}) \cap R(\bX_{1DO2})=\{0\}$ and $R(\bX_{1DO23}) \cap R(\bX_{1DO3})=\{0\}$. Hence, the full decomposition of $R(\bX_1)$ becomes\\
\begin{gather}\label{eDirectSum4}
    R(\bX_1) = R(\bX_{123C}) \oplus R(\bX_{12C}) \oplus R(\bX_{13C}) \oplus R(\bX_{1DO23}) \\ \nonumber \oplus R(\bX_{1DO2}) \oplus R(\bX_{1DO3}) \oplus R(\bX_{1DNO})
\end{gather}

\noindent that represents the most elaborate decomposition of $R(\bX_1)$ if all dimensions allow so with different possibilities for orthogonalities. Because of the direct sum properties the dimensions add up in the same way as in Eq. \ref{eDim1} and \ref{eDim2}. Similar decompositions can be made for $R(\bX_2)$ and $R(\bX_3)$.  Schematically, the decomposition of Eqn. \ref{eDirectSum4} is shown in Figure \ref{Fig_ThreeBlocks}.

\noindent Equations \ref{eDirectSum2}, \ref{eDirectSum3} and \ref{eDirectSum4} show an increasing degree of complexity. We give here the full decompositions to be complete, but it is important to mention that in most practical cases, one is not interested in all these subspaces making the actual practical decomposition simpler. This is even more so in cases with more than three blocks.

\subsection{Theoretical considerations}
\label{Theoretical considerations}

In practice, data always contain noise and therefore we cannot expect to find a decomposition that satisfies all the idealistic requirements described above. Also, which decomposition to make under which constraints depends very much on the type of application.  Before showing how various types of already existing methods try to solve this challenge, we will here discuss some of the major issues that have to be taken into account. These issues represent choices which have to be made regarding the nature of the common and distinct components, diagnostic tools such as explained sum-of-squares, scaling of the variables and the data sets.

\subsubsection{Fundamentally different choices of common components.}

Of particular importance here is the concept of common variation because it can be considered as a starting point of the decomposition. Since practical implementations are usually based on extracting components or basis vectors for the different spaces, most of the following discussions will be related to components rather than to general vector spaces as was the case above.\\

\begin{figure}[h!]
 \centerline{\includegraphics*[width=14cm]{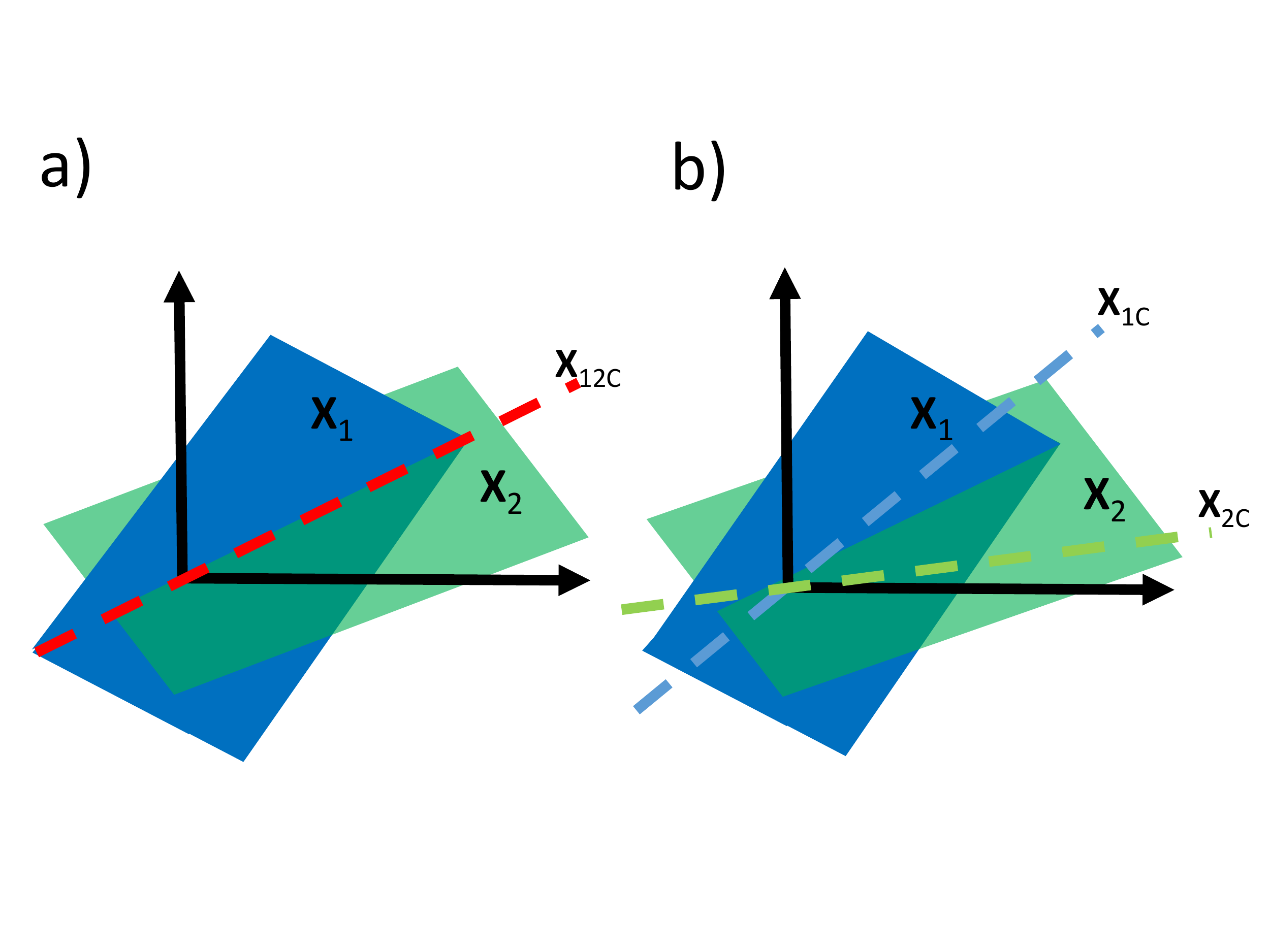}}
 \caption{\footnotesize The common subspace under noisy conditions. It can be chosen as a compromise in-between $R(\bX_1)$ and $R(\bX_2)$ but not a part of neither of those (red dashed line, a) or as parts of their respective column-spaces but unequal to each other (blue and green lines, b).}
 \label{Fig_Common_Noise}
\end{figure}

\noindent In noisy data, the situation as shown in Figure \ref{Fig_Common} does not usually hold: there is no common space in mathematical terms (an intersection) because the column-spaces have changed due to the noise. There are two fundamentally different categories of approaches and these are present in the methods that are discussed in Section \ref{How established methods relate to the definitions}. In the first category, a common component is found as the best compromise solution between the two column-spaces: vector $\bX_{12C}$ in Figure \ref{Fig_Common_Noise}a (although it is customary to use a bold-lowercase character for a vector, we keep the notation using a matrix to stress the fact that we are generally discussing subspaces). This vector is neither in the column-space of $\bX_1$ nor in the column-space of $\bX_2$. In the second category, a different choice is made. The common component is estimated separately in each column-space. Hence, rather than one common component, two separate ones are found but generally in a manner that seeks them to be as similar as possible(although $\bX_{1C} \neq \bX_{2C}$; Figure \ref{Fig_Common_Noise}b) and thus they can be seen as representing a common component. Both choices are made in the methods to be discussed and both approaches have their pros and cons (see Table \ref{Table3} for more details). Note the change in notation of the common parts to emphasize this difference.

\subsubsection{Sums of Squares and Explained Variation.}

When it comes to assessing the importance of a subspace in the decomposition there are at least two aspects that have to be taken into account; the dimension of the subspaces identified and variances explained by those subspaces in the original data. The former relates to how many linearly independent components are estimated to form the subspace. The latter relates to the contributions of the subspaces to the total variation in a block. If orthogonality is used in the decomposition when defining the distinct space (Figure \ref{Fig_Distinct}a), it is easy to show that the total sum of squares (SS) for a block can be split in one contribution from the common space ($R(\bX_{12C})$) and one for the orthogonal distinct contribution ($R(\bX_{1D})$):
\begin{equation}\label{eSS1}
    \|\bX_1\|^2=\|\bX_{12C}\|^2+\|\bX_{1D}\|^2
\end{equation}
where we use the symbol $\|.\|^2$ to indicate the squared Frobenius norm of a matrix. An analogous equation can be written for $\bX_2$. If orthogonality is not imposed between the common and distinct parts (Figure \ref{Fig_Distinct}b), a decomposition of SS is still possible, but the interpretation of the last term is different. In that case, it is simply defined as the additional variation that is explained by adding the distinct components, i.e. as $\|\bX_1\|^2=\|\bX_{12C}\|^2+\|\widetilde{\bX}_{1D}\|^2$, where $\widetilde{\bX}_{1D}$ is the part of $R(\bX_{1D})$ orthogonal to $R(\bX_{12C})$. This is sometimes called Extra Sum of Squares (ESS; see also \cite{Peres-Neto2006}). Note that the order in which the terms are calculated in Eqn. \ref{eSS1} matters in the non-orthogonal case. For the orthogonal case, the two interpretations coincide.\\

\noindent When decomposing the distinct part further into an orthogonal and a non-orthogonal part, the resulting (E)SS can be written as
\begin{equation}\label{eSS2}
    \|\bX_1\|^2=\|\bX_{12C}\|^2+\|\bX_{1DO}\|^2+\|\bX_{1DNO}\|^2
\end{equation}
and the interpretation depends on the orthogonality properties. In the most extreme case, all subspaces $R(\bX_{12C})$, $R(\bX_{1DO})$ and $R(\bX_{1DNO})$ are orthogonal to each other, then Eqn. \ref{eSS2} can be interpreted in terms of sums of squares of contributions of each block. In all other cases, Eqn. \ref{eSS2} has an ESS interpretation, e.g., when $R(\bX_{1DO})$ and $R(\bX_{1DNO})$ are not orthogonal, then $\|\widetilde{\bX}_{1DNO}\|^2$ is the ESS of the distinct-non-orthogonal part where $R(\widetilde{\bX}_{1DNO})$ is orthogonalized relative to $R(\bX_{1DO})$. Explained variation of common or distinct components within a block can now be calculated and expressed as percentages of the total variation in that block. Note that this process is analogous to the Type I ANOVA where focus is on additional contribution of variables in explaining a Y-variable and note also the similarity to SO-PLS in a multi-block regression context \cite{Naes2013}.\\

\begin{figure}[h!]
 \centerline{\includegraphics*[width=14cm]{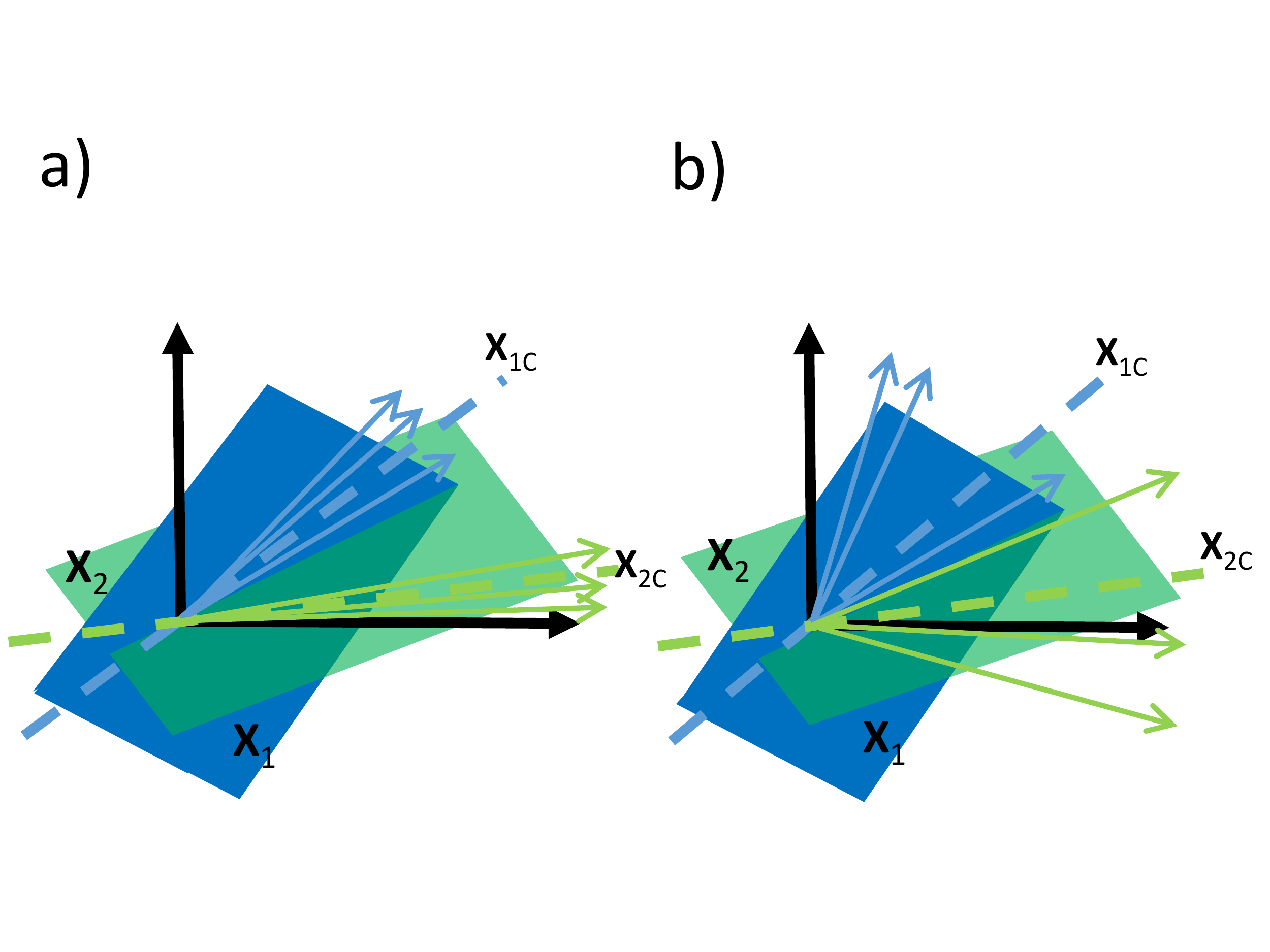}}
 \caption{\footnotesize Well-embedded  common component (a) and poorly embedded common direction (b).}
 \label{Fig_Embed}
\end{figure}

\noindent The issue of variance explained by common components in a block is visualized in Figure \ref{Fig_Embed} for the second category of methods. The column vectors making up the column-spaces of $\bX_1$ and $\bX_2$ are explicitly drawn in the figure. In the left (a) part all these column vectors (i.e. variables) are close to the common components within each block. Hence, the common components are representative of their respective column-spaces: they are embedded well and explain a high amount of variation in each block. This is not the case for the right (b) part of the figure: the common component $\bX_{1C}$ is not well embedded in $\bX_1$. Usually, explaining within-block variation and having between-block correlation cannot be achieved simultaneously and a good account of this trade-off is given elsewhere \cite{VandeGeer1984}.

\subsubsection{High-dimensional data.}

High-dimensional data need some extra considerations. This type of data is abundant in modern scientific fields such as genomics, e.g., when considering gene-expression data where the number of genes (variables) is much larger than the number of samples. In our framework, there is now \textit{necessarily} a common subspace simply due to the dimensions. For instance, if $\bX_1$ has size $ 20 \times 1000$ and $\bX_2$ has size $ 20 \times 10000$ both of rank 20, then they trivially share the same 20-dimensional column-space which is thus $R(\bX_{12C})$. In such cases, calculating for instance canonical correlations is problematic and some type of regularization is necessary. Without such regularization, the chances are in most cases high that only uninteresting, trivial and noisy components are identified. One way of trying to solve the problem with many variables and few objects is to use PCA for each block separately, in this way reducing both the noise and the dimensionality (see \cite{VandenBerg2009a,Mage2012}). Whether this approach is preferable depends on a number of properties (ranks of the different subspaces, noise characteristics etc.) and other approaches will be discussed later in this paper.

\subsubsection{Scaling issues.}

Another general aspect that is important to discuss is the scaling of the blocks and variables within blocks. We will refer to this as between-block scaling and within-block scaling. One example of the latter is known as auto-scaling in chemometrics, standardization in psychometrics and normalization in statistics. We assumed already centered data and auto-scaling on top of that also divides every column of a matrix by its standard deviation. Hence, the data is analyzed in correlation mode. The between-block scaling is related to the total variation of a block. It is often natural to do some type of overall scaling of the blocks in order to avoid too much dominance of one of the blocks. For instance, in cases where one of the blocks has only a few variables and another one has many variables, the joint approach could put almost all emphasis on trying to model the larger data set. This may lead to solutions where one is not modeling the joint variation, but merely within block variability, which is clearly not the intention in data fusion. A possible way to counter this is to divide each block by the Frobenius norm prior to analysis. General guidelines for centering and scaling are available \cite{Bro2003,VandenBerg2006} and there is also literature on scaling in multi-block data analysis \cite{VanDeun2009a,Simsekli2013,Wilderjans2011,Timmerman2015}.

\section{How established methods relate to the definitions}
\label{How established methods relate to the definitions}

In this section, we will discuss how a number of already existing methods aiming for identifying common and distinct components are related to the definitions given in Section 2. We will discuss these methods mostly by using two blocks of data and more than two blocks if clarity allows to do so (we will index the blocks by $k=1,...,K$). Tables \ref{Table1} and \ref{Table2} summarize properties of these discussed methods. The methods to be discussed originate from different fields of science and thus use different notations. We will try to harmonize this by using as much as possible a uniform notation based on the familiar PCA model:
\begin{equation}\label{ePCA}
    \bX = \bX \bW \bP^T+ \bE = \bT \bP^T + \bE
\end{equation}
where the matrix of weights $\bW$ defines linear combinations of the columns of $\bX$, generating scores $\bT$ and loadings $\bP$ which are the regression coefficients of $\bX$ on $\bT$. In PCA, the matrix $\bW$ will be identical to $\bP$, but this is not necessarily so for all methods. To arrive at a consistent terminology for all the methods to be discussed, we will use the terms and corresponding symbols weights, scores and loadings in the following.

\subsection{Simultaneous Component Analysis, Generalized Canonical Correlation and a compromise}

The two most different ways of defining common variability are probably PCA on the concatenated matrix $[\bX_1|...| \bX_K]$ which focuses on explaining the simultaneous variation in all blocks and Canonical Correlation or its generalized form (GCA; see below) which only focuses on correlation between the blocks. The PCA model on concatenated data goes under various names as will be explained in the next section.

\subsubsection{Simultaneous Component Analysis}

The optimization criterion for Simultaneous Component Analysis (SCA) is
\begin{equation}\label{eSCA1}
    \min_{(\bT,\bP_k)}\sum_{k=1}^K \|\bX_k-\bT \bP^T_k\|^2
\end{equation}
where the simultaneous components are represented by $\bT (I \times R)$ and the loadings $\bP_k (I \times J_k)$ measure how these components are related to the original data. This model is known under different names: SUM-PCA in chemometrics \cite{Smilde2003}, Simultaneous Component Analysis (SCA-P) in psychometrics \cite{Timmerman2003} and Tucker1 in three-way analysis \cite{Smilde2004}. The underlying idea of using this model is that $\bT$ represents as much as possible the variation in all data blocks simultaneously. Hence, a model of each block can be written as
\begin{equation}\label{eSCA2}
    \bX_k=\bT \bP^T_k+\bE_k;k=1,...,K
\end{equation}
and several properties of SCA are described in Tables \ref{Table1} and \ref{Table2}. The optimization problem of Eqn. \ref{eSCA1} is stated as a least-squares problem, but can also be formulated as the problem of finding the eigenvectors of
\begin{equation}\label{eSCA3}
    \bZ_{SCA}=\sum_{k=1}^K\bX_k \bX^T_k
\end{equation}
and selecting the $R$ eigenvectors belonging to the $R$ largest eigenvalues. Alternatively, the components $\bT$ can be found using the SVD of $[\bX_1|..|\bX_K]$ and choosing the $R$ left singular vectors corresponding to the $R$ largest singular values (i.e. a PCA on the concatenated matrix $[\bX_1|...| \bX_K]$). Hence, the components $\bT$ are in the column-space of $[\bX_1|..|\bX_K]$ and not necessarily in the column-spaces of any of the individual matrices $\bX_k$. The matrix $\bT$ represents both common and distinct variation according to the definitions given above and, hence, the model is not separating common and distinct sources of variation. Moreover, the term \emph{simultaneous} component analysis suggest a focus on common components which is not the case. Nevertheless, we present SCA here since it is much used in multi-set analysis and a starting point of other methods. Note that the least squares property does not hold per data block, but only across all blocks simultaneously. However, given $\bT$, Eqn. \ref{eSCA2} is a least squares model for the set of all $\bX_k$.\\

\noindent Without loss of generality, the simultaneous components $\bT$ can be chosen to be orthogonal due to rotational freedom of the model. The subspace spanned by $\bT$ is unique like in ordinary PCA. The residuals $\bE_k$ are orthogonal to the model part of $\bX_k$ (which is $\bT \bP^T_k$) and thus a break-down of sum-of-squares can be calculated. Note, however, that due to the fact that $\bT$ is not necessarily in the range of $\bX_k$ neither is $\bE_k$. SCA is sensitive to between- and within-block scaling.\\

\noindent Whereas SCA is a simultaneous method for data fusion, there is a history of sequential methods in chemometrics which serve the same purpose. These methods are known under different names and versions (Hierarchical PCA, Consensus PCA, Multiblock PCA). Due to their sequential nature, it is sometimes difficult to assess their properties, but some results exist \cite{Westerhuis1998,Smilde2003}.\\

\noindent SCA has been used in several areas of science and is a special case of a much broader method in data mining called Collective Matrix Factorization \cite{Singh2008}. In metabolomics and process chemometrics it is used in conjunction with multilevel data analysis and as a step after an initial ANOVA \cite{Smilde2005a,Jansen2005a,Noord2005}. It is also used in spectroscopy \cite{Bevilacqua2013,Tao2013,Tomassini2014,Shan2015} and in sensory science \cite{Pages2005,Dahl2006,Bro2008}.

\subsubsection{Generalized Canonical Correlation Analysis (GCA)}
\label{Generalized Canonical Correlation Analysis (GCA)}

The goal of GCA is to identify linear combinations of the blocks, $\bX_k \bW_k$, which fit as well as possible to a set of orthogonal common components $\bT$. This is done by minimizing the criterion
\begin{equation}\label{eGCA1}
    \min_{(\bT,\bW_k)} \sum^K_{k=1}\|\bX_k \bW_k-\bT\|^2
\end{equation}
with respect to $\bT  (\bT^T \bT=\bI)$ and $\bW_k (k=1,...,K)$ \cite{VanderBurg1996}. The number of columns in $\bT$, $A$, must be smaller than or equal to the number of columns in the $\bX_k$ with the smallest number of columns. If the number of samples, $I$, is smaller than all $J_k (k=1,...,K)$, then $A=I$ is the maximum number of components. Note that the same solution can be obtained by maximizing a sum of correlations between linear combinations of the X blocks, which is the typical formulation for the situation with only two X-blocks \cite{Hotelling1936}. In that case, this is usually referred to as canonical correlation analysis. In practice, the actual solution $\bT$ is found as the eigenvectors of the matrix
\begin{equation}\label{eGCA2}
    \bZ_{GCA} = \sum^K_{k=1} \bX_k (\bX^T_k \bX_k)^+ \bX^T_k
\end{equation}		
where again the + means the Moore-Penrose (pseudo-)inverse. The $\bW_k's$ can then be found by regressing $\bT$ on $\bX_k$: $\bW_k = \bX^+_k \bT$.\\

\noindent If there are $A$ common components in the X-blocks according to the definition given in Section \ref{Description of the framework}, the criterion in Eqn. \ref{eGCA1} will exactly be equal to 0. In the two-block case the common components will correspond to components with a canonical correlation equal to one.\\

\noindent The solution $\bT$ in Eqn. \ref{eGCA1} is not necessarily within any of the column-spaces of the $\bX_k's$ but it is in the column space of $[\bX_1|...|\bX_K]$ (for a proof, see the Appendix). Although this is not the goal of GCA, when needed a model of $\bX_k$ can be obtained by regressing $\bX_k$ on $\bT_k = \bX_k \bW_k$ giving loadings $\bP_k$ from which also explained variances can be calculated(see Table \ref{Table1}).\\

\noindent Since GCA only concentrates on correlation and gives no emphasis on within block variability (thereby potentially poorly embedded and hence unstable), several methods have been developed for balancing the two aspects. One particular solution is obtained by defining a continuum of solutions between SCA and GCA using a ridge regression type of formulation joining Eqns. \ref{eSCA3} and \ref{eGCA2} in one single formula \cite{Dahl2006}. Enhancing stability of the GCA components can also be obtained by using PCA on the individual data blocks data before using GCA \cite{VandenBerg2009a} or by regularization \cite{Tenenhaus2014}. The solution using PCA as a first step will be called PCA-GCA in the example section.\\

\noindent It is possible to also obtain distinct components using PCA-GCA. This can be done by regressing each block on its own common components. The residuals from these regressions represent two distinct subspaces $R(\bX_{1D})$ and $R(\bX_{2D})$ which can subsequently be subjected to a PCA for each subspace. Note that in this case $R(\bX_{1D})$ is orthogonal to $R(\bX_{1C})$ and likewise $R(\bX_{2D})$ is orthogonal to $R(\bX_{2C})$, but $R(\bX_{1D})$ is not necessarily orthogonal to $R(\bX_{2D})$. Hence, we are in the situation of Figure \ref{Fig_Distinct}a. GCA does not depend on within-block and between-block scaling and thus the distinct subspaces also do not depend on that. However, performing a PCA on the distinct subspaces depends of course on the within-scaling of the distinctive matrices. Examples of the use of GCA can be found, e.g., in sensory science \cite{Dahl2006,VanderBurg1996}. Also in signal processing GCA-type methods are used \cite{Correa2010} based on the work in biometrics \cite{Kettenring1971}.\\

\subsection{O2PLS}

There seem to be three different implementations of O{\footnotesize 2}PLS \cite{Trygg2002,Trygg2003,Lofstedt2011}. The last implementation is a generalization of O{\footnotesize 2}PLS to OnPLS (for more than two blocks). The O{\footnotesize 2}(n)PLS methods are usually described in terms of iterative algorithms rather than through formal definitions of well-defined criteria which makes their properties difficult to assess. We describe the implementation of Lofstedt \citep{Lofstedt2012}.

The starting point for O{\footnotesize 2}PLS is the SVD of the covariance matrix $\bX^T_2 \bX_1$:			\begin{equation}\label{eO2PLS1} 										
 		\bU \bS \bV^T = \bX^T_2 \bX_1								
\end{equation}
and collecting the $R$ singular vectors corresponding to the $R$ largest singular values of Eqn. \ref{eO2PLS1} in $\bU_R$ (left-singular vectors) and $\bV_R$ (right-singular vectors), respectively, as weights for the (preliminary) common components. This SVD is known as the product SVD (PSVD) and is a member of a broad class of generalizations of the ordinary SVD \cite{DeMoor1991,DeMoor1992} and Eqn. \ref{eO2PLS1} is actually also the first step of Bookstein's version of PLS \cite{Bookstein1994}. Define $\bF_1=\bX_1-\bX_1\bV_R\bV^T_R$ and $\bF_2=\bX_2-\bX_2\bU_R\bU^T_R$ then due to the truncation to $R$ components, the (preliminary) common components $\widetilde{\bT}_{1C} = \bX_1 \bV_R$ still share some variation with $\bF_1$ and likewise $\bX_2 \bU_R$ with $\bF_2$. This part can be calculated by solving
\begin{equation}\label{eO2PLS2} 										
 		\max_{\|\bz_1\|=1} \|\widetilde{\bT}^T_{1C} \bF_1 \bz_1 \|^2 								
\end{equation}
which maximizes the shared variation of $\widetilde{\bT}_{1C}$ and $\bF_1$ in one (orthogonal) component (indexed by $l=1,...,L$). A deflation procedure then subsequently regresses $\bX_1$ on this component $\bF_1 \bz_1$ and gives the residuals $\bX_{res1}$. A similar procedure can be used for $\bX_2$, and the number of orthogonal components has to be chosen (or found). Then (final) common components between the deflated matrices can be extracted one by one by using the MAXDIFF criterion \cite{Hanafi2006}:
\begin{equation}\label{eO2PLS3} 										
 		\max_{\bw_1,\bw_2} tr(\bw^T_1 \bX^T_{res2} \bX_{res1} \bw_2) = tr(\bt^T_{2C} \bt_{1C});\bW^T_k \bW_k=\bI (k=1,2)								
\end{equation}
where the matrices $\bT_{1C}$ and $\bT_{2C}$ collect the vectors $\bt_{1C}$ and $\bt_{2C}$, respectively, and the matrix $\bW_k; k=1,2$ contain the weights for the different dimensions. This will result in the following models for $\bX_1$ and $\bX_2$:
\begin{eqnarray}\label{eO2PLS4}
    \bX_1 &=& \bT_{1C} \bW^T_1 + \bT_{1D} \bP^T_{1D} + \bE_1 = \bX_{1C} + \bX_{1D} + \bE_1 \\ \nonumber
    \bX_2 &=& \bT_{2C} \bW^T_2 + \bT_{2D} \bP^T_{2D} + \bE_2 = \bX_{2C} + \bX_{2D} + \bE_2
\end{eqnarray} 						
where $\bT_{1D}$ collects the orthogonal components $\bF_1 \bz_l$ (hence the name O(rthogonal){\footnotesize 2}PLS) and $\bP_{1D}$ its loadings, and likewise for $\bT_{2D}$ and $\bP_{2D}$. The orthogonality properties between the different matrices are shown in Table \ref{Table2} (see \cite{VanderKloet2015}). This means that O{\footnotesize 2}PLS takes the viewpoint of Figure \ref{Fig_Distinct}a (apart from the fact that each block has its own common component). Eqn. \ref{eO2PLS4} shows that also O{\footnotesize 2}PLS fits in our framework but calculating explained variances is hampered by the orthogonality properties. Note again that we changed notation of the common parts to emphasize that $R(\bX_{1C}) \neq R(\bX_{2C})$. O{\footnotesize 2}PLS is within-block scale dependent but between-block scale independent.\\

\noindent The O{\footnotesize 2}PLS method has been used amongst others in spectroscopy \cite{Mattarucchi2010,Consonni2010,Kirwan2013,Petrakis2015}, in the plant sciences \cite{Bylesjo2007,Szymanski2014} and its extension to more than two blocks (OnPLS) has been used in genomics \cite{Srivastava2013,Lofstedt2013}. The latter paper also describes an implementation of the multi-block problem as shown in Figure \ref{Fig_ThreeBlocks} showing the complexity of such a decomposition. There is an interesting relationship of O{\footnotesize 2}PLS with Procrustes analysis, as explained in the Appendix.

\subsection{DIStinct and COmmon-Simultaneous Component Analysis (DISCO-SCA)}
\label{DISCO}
Also the DISCO-SCA (or DISCO, for short) method \cite{Schouteden2013,VanDeun2012} can be posed in terms of our framework. The first step in DISCO is to solve an SCA problem to find scores $\widetilde{\bT} (I \times R)$ and loadings $\widetilde{\bP} ((J_1+J_2) \times R)$ of the concatenated matrix $[\bX_1|\bX_2]$. The loading matrix $\widetilde{\bP}$ can be partitioned in $\widetilde{\bP}_1 (J_1 \times R)$ and $\widetilde{\bP}_2 (J_2 \times R)$.  Subsequently, the matrix $\widetilde{\bP}$ is orthogonally rotated to a simple structure reflecting distinct and common components. For the sake of illustration, assume that $R=3$; there are one common and two distinct components (one for each block). Then $\widetilde{\bP}$ is orthogonally rotated to a structure $\bP_{target}$ according to
\begin{equation}\label{eDISCO1} 										
 		\min_{\bQ^T \bQ=\bI} \left\|\bV \odot (\widetilde{\bP} \bQ - \bP_{target})\right\|^2						
\end{equation}
with $\bV$ a matrix of zero's and one's selecting the elements across which the minimization occurs, the symbol $\odot$ indicates the Hadamard or elementwise product and
\begin{equation}\label{eDISCO2}
    \bP_{target} = \left| \begin{array}{ccc} x & 0 & x \\ x & 0 & x \\ x & 0 & x \\ 0 & x & x \\ 0 & x & x \\ 0 & x & x \\ 0 & x & x \\ \end{array} \right|
\end{equation}
where the symbol $x$ means an arbitrary value not necessarily zero and $\bP=\widetilde{\bP}\bQ=[\bP^T_1 | \bP^T_2]^T$. This will result in the first component being distinct for $\bX_1$, the second component distinct for $\bX_2$ and the third component will be the common one. After finding the optimal $\bQ$, the scores $\widetilde{\bT}$ are counter-rotated resulting in $\bT=\widetilde{\bT} \bQ = [\bt_1 \bt_2 \bt_3 ]$ and the following decomposition is obtained:
\begin{eqnarray}\label{eDISCO3}
    \bX_1 &=& \bT \bP^T_1 = \bt_1 \bp^T_{11} + \bt_2 \bp^T_{12} + \bt_3 \bp^T_{13} + \bE_1 \\ \nonumber
    \bX_2 &=& \bT \bP^T_2 = \bt_1 \bp^T_{21} + \bt_2 \bp^T_{22} + \bt_3 \bp^T_{23} + \bE_2
\end{eqnarray}
where $\bp_{11}$ gives loadings for the distinct component for $\bX_1$; $\bp_{22}$ for the distinct component for $\bX_2$ and $\bp_{13}, \bp_{23} $ for the common component. If $\bp_{12}$ is not close to zero then there is a distinct non-orthogonal part in the decomposition of $\bX_1$ (the red colored $\bX_{1DNO}$ in Table \ref{Table1}). The SCA solution has orthogonal columns in $\widetilde{\bT}$ and rotates orthogonally afterwards thus these columns remain orthogonal. Hence, $\bT$ is orthogonal and it holds that $\bX_{1DNO}$ is orthogonal to both $\bX_{1DO}$ and $\bX_{1C}$, but it is clearly not orthogonal to $\bX_{2DO}=\bt_2 \bp^T_{22}$ (see Table \ref{Table3} and Figure \ref{Fig_Distinct}a). Minimizing the sum of squared elements of $\bp_{12}$ and $\bp_{21}$ is exactly what the above mentioned rotation tries to do, thereby minimizing the sizes of these distinct non-orthogonal parts and defining those parts as being distinct non-orthogonal (see the $\bP_{target}$ in Eqn. \ref{eDISCO2}). Thus, this is yet another implementation of the general decomposition scheme where the vectors can be matrices when more than one common and distinct components are present. Contrary to O2PLS, the common parts in DISCO span the same column-space. Because DISCO starts with an SCA and subsequently utilizes a rotation, both $\bT$ and $\widetilde{\bT}$ are in the column-space of the combined $[\bX_1 | \bX_2]$ rather than the individual parts. The uniqueness properties of DISCO are unknown. Due to the orthogonality of the scores matrix $\bT$ explained variances can be calculated based on Eqn. \ref{eDISCO3}. DISCO is within- and between-block scale dependent and has been used in metabolomics \cite{VanDeun2012} and in gene-expression analysis \cite{VanDeun2013}.

\subsection{Generalized Singular Value Decomposition (GSVD)}
\label{GSVD}
A method used in gene-expression data to separate common from distinct components is the Generalized Singular Value Decomposition (GSVD) \cite{Alter2003} which is also a generalization of the SVD known as the Quotient SVD (QSVD) \cite{DeMoor1991}. The mathematics of the GSVD dates back already some time \cite{VanLoan1976,Paige1981}. The original GSVD is used for fusion of data sharing the same columns but this problem can be transposed to our situation. The original GSVD is a matrix decomposition method and does not have least squares properties. To repair its sensitivity to noise, we follow the implementation of the Adapted GSVD which comes down to first filtering the data with an SCA step \cite{VanDeun2012}. For the two-block case the model is
\begin{eqnarray}\label{eGSVD1}
    \bX_1 &=& \widehat{\bX}_1 + \bE_1 = \bT \bD_1 \bV^T_1 + \bE_1 \\ \nonumber
    \bX_2 &=& \widehat{\bX}_2 + \bE_2 = \bT \bD_2 \bV^T_2 + \bE_2
\end{eqnarray}
with $\widehat{\bX}_k$ is the filtered data; $\bV^T_k \bV_k = \bI (k=1,2)$, $\bD_k (k=1,2)$ diagonal and such that $\bD^2_1 + \bD^2_2 = \bI$, and $\bT$ a full-rank matrix but not necessarily orthogonal. Due to the latter constraint it is possible to divide the generalized singular values (the elements of $\bD_k (k=1,2)$)  in three groups: if $d^2_{1R} \approx 1$ the corresponding component is distinctive for $\bX_1$, if $d^2_{2R} \approx 1$ the corresponding component is distinctive for $\bX_2$ and if $d^2_{1R} \approx d^2_{2R}$ the corresponding component is common. Obviously, there is a certain amount of arbitrariness in these choices. Once such a choice is made, Eqn. \ref{eGSVD1} can be written as
\begin{eqnarray}\label{eGSVD2}
    \bX_1 = \bT_1 \bD_{11} \bV^T_{11} + \bT_2 \bD_{12} \bV^T_{12} + \bT_3 \bD_{13} \bV^T_{13} + \bE_1 \\ \nonumber
    \bX_2 = \bT_1 \bD_{21} \bV^T_{21} + \bT_2 \bD_{22} \bV^T_{22} + \bT_3 \bD_{23} \bV^T_{23} +\bE_2
\end{eqnarray}
which fits our framework. Upon assuming that $\bT_1 \bD_{11} \bV^T_{11}=\bX_{1DO}$, $\bT_2 \bD_{22} \bV^T_{22}=\bX_{2DO}$ and $\bT_3 \bD_{13} \bV^T_{13},\bT_3 \bD_{23} \bV^T_{23}$ are the common components, then $\bT_2 \bD_{12} \bV^T_{12}=\bX_{1DNO}$ and $\bT_1 \bD_{21} \bV^T_{21}=\bX_{2DNO}$. Due to the orthogonality of both $\bV_1$ and $\bV_2$ it holds that $\bX_{1DO}$, $\bX_{1DNO}$ and $\bX_{1C}$ are mutually orthogonal, and likewise for block $\bX_2$. However, $\bX_{1DNO}$ is not orthogonal to $\bX_{2DO}$ and similarly $\bX_{2DNO}$ is not orthogonal to $\bX_{1DO}$. This is the same as for DISCO and is again the situation of Figure \ref{Fig_Distinct}a.\\

\noindent For an invertible $\bX_2$ the GSVD equals the SVD of $\bX_1 \bX^{-1}_2$ which explains the term Quotient SVD. For these cases, the uniqueness properties of the GSVD are the same as those of the SVD. For non-invertible $\bX_2$ the uniqueness properties are not clear. GSVD is within- and between-block scale dependent and has been used in gene-expression analysis \cite{Alter2003} and has been extended for more than two blocks in different ways \cite{DeLathauwer2009,Ponnapalli2011}.

\subsection{Joint and Individual Variances Explained (JIVE)}
\label{JIVE}

The method of Joint and Individual Variances Explained (JIVE \cite{Lock2013}) goes as follows. For two blocks, it derives directly a decomposition according to:
\begin{eqnarray}\label{eJIVE1}
    \bX_1 &=& \bT_C \bP^T_{1C} + \bT_{1D} \bP^T_{1D} + \bE_1 = \bX_{1C} + \bX_{1D} + \bE_1 \\ \nonumber
    \bX_2 &=& \bT_C \bP^T_{2C} + \bT_{2D} \bP^T_{2D} + \bE_2 = \bX_{2C} + \bX_{2D} + \bE_2
\end{eqnarray}
which fits in our framework. Note that we use the notation $\bT_C$ to stress that the common scores are the same. In estimating this decomposition, the following constraints are used
\begin{equation}\label{eJIVE2}
    \bX_{1C}^T \bX_{1D} = 0; \bX_{1C}^T \bX_{2D} = 0; \bX_{2C}^T \bX_{1D} = 0; \bX_{2C}^T \bX_{2D} = 0
\end{equation}
and, thus, the distinct part in a block is orthogonal to the common parts in all blocks but the distinct parts in different blocks are not necessarily orthogonal. This is again an implementation of Figure \ref{Fig_Distinct}a. The (low) ranks of all common and distinct matrices involved are determined by permutation tests. Since $\bE_k$ is not necessarily orthogonal to neither $\bX_{kC}$ nor $\bX_{kD}$, separating sums-of-squares (and variances, despite the name) is not easy for JIVE. For other properties, see Tables \ref{Table1} and \ref{Table2}. JIVE is within- and between-block scale dependent and has been applied in gene-expression analysis \cite{Lock2013}.

\subsection{Structure Revealing Data Fusion}
\label{SR}

In Structure Revealing Data Fusion \cite{Acar2014} an approach is chosen based on penalties. The method is developed for fusing two-way and three-way arrays but can equally well be used for fusing two-way arrays. Starting point is the model
\begin{eqnarray}\label{eSRDF1}
    \bX_1 &=& \bT \bD_1 \bV^T_1 + \bE_1 = \bT \bP^T_1 + \bE_1  \\ \nonumber
    \bX_2 &=& \bT \bD_2 \bV^T_2 + \bE_2 = \bT \bP^T_2 + \bE_2
\end{eqnarray}
where the matrices $\bD_1$ and $\bD_2$ are diagonal and the diagonals of $\bT^T \bT$, $\bV^T_1 \bV_1$ and $\bV^T_2 \bV_2$ consist of ones (i.e. the columns of $\bT$, $\bV_1$ and $\bV_2$ have length one). The components are now estimated under an $L_1$ penalty \cite{Tibshirani2011}:
\begin{equation}\label{eSRDF2} 										
 		\min_{\bT,\bV_1,\bV_2,\bD_1, \bD_2} \left\|\bX_1-\bT \bD_1 \bV^T_1\right\|^2+\left\|\bX_2-\bT \bD_2 \bV^T_2\right\|^2+\lambda (\left\|diag(\bD_1)\right\|_1 + \left\|diag(\bD_2)\right\|_1) 						
\end{equation}
where $\lambda \geq 0$ is the penalty parameter to be set by the user, the symbol $\left\|.\right\|_1$ represents the $L_1$-norm and $diag(\bD_k)$ is the vector carrying the diagonal of $\bD_k$. Increasing the penalty value $\lambda \geq 0$ will force more elements in $\bD_1$ and $\bD_2$ to become zero. From the patterns of these zero's the common and distinct components are defined. This type of approach - albeit in fusing three-way and two-way data - has been used in metabolomics \cite{Acar2014,Acar2015}. Structure Revealing Data Fusion is within- and between-block scale dependent.\\

\noindent A special class of Structure Revealing Data Fusion methods are the multivariate curve resolution methods as used in chemometrics \cite{Tauler1995}. This class of methods performs data fusion mostly using hard constraints on the parameters based on chemical information. There are very many applications of this method in different fields of chemistry.

\section{Examples}
\label{Examples}

\noindent To illustrate some of the methods falling under our framework and their relationships, we will show some real data examples that were already introduced shortly in Section \ref{Introduction and Motivation}. Two real data sets from medical biology and food science will be used to show aspects related to practical use of the methods. All data will be analyzed by the methods PCA-GCA (see Section \ref{Generalized Canonical Correlation Analysis (GCA)}) and DISCO (see Section \ref{DISCO}). These methods are selected because they represent different orthogonality constraints (referring to Figure \ref{Fig_Distinct}a) and b), respectively). They also represent different choices of common components as discussed in Section \ref{Theoretical considerations}: PCA-GCA estimates separate common components for each data block while DISCO estimates a “best compromise” which is in the column space of the concatenated data blocks.

For all methods, the first step is to decide the dimensionalities of the subspaces. This is not a trivial task and different strategies exist for the different methods. The strategies for PCA-GCA and DISCO will be explained briefly in the examples below, but a thorough discussion of the model selection is not within the scope of this paper. Once the dimensions are decided, it is straightforward to estimate basis vectors (or components) for each of the subspaces.

\subsection{Sensory example}

\noindent The sensory example focuses on one of the typical aspects of a product development process: The product developer is interested in understanding how well two important modalities of the descriptive sensory profile relate to the ingredients in the recipe. A typical issue of interest for being able to optimize product quality is whether the recipe influences both smell and taste and in which way this happens. In particular, one is interested in knowing what aspects of smell and taste that are common and what is unique in the two sensory profiles.\\

\noindent This example consists of descriptive sensory attributes of flavored water samples and is a subset of a larger data set \cite{Mage2012}. The 18 water samples are created according to a full factorial experimental design with two flavor types (A and B), three flavor doses (0.2, 0.5 and 0.8 g/l) and three sugar levels (20, 40 and 60 g/l). A trained sensory panel consisting of 11 assessors evaluated the samples first by smelling (9 descriptors) and then by tasting (14 descriptors), using an intensity scale from 1 to 9. Two data blocks (SMELL and TASTE) were constructed by averaging across assessors. The blocks were mean-centered and block-scaled to sum-of-squares one prior to analysis.\\

\begin{figure}[h!]
 \centerline{\includegraphics*[width=14cm]{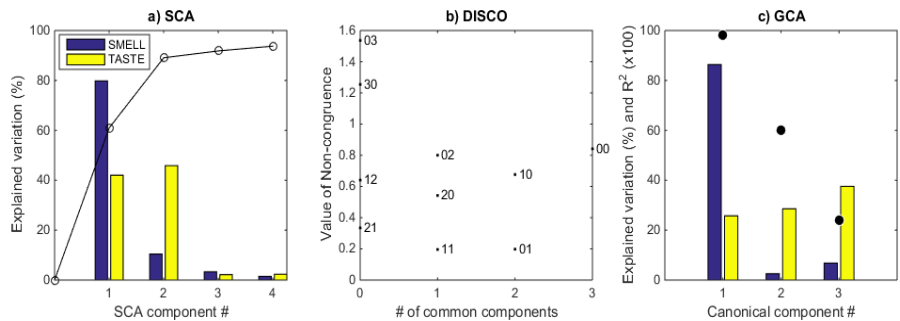}}
 \caption{\footnotesize Plots for selecting numbers of components for the sensory example. a) SCA. The curve represents cumulative explained variance for the concatenated data blocks. The bars show how
much variance each component explain in the individual blocks. b) DISCO. Each point represent the non-congruence value for a given target (model). The plot includes all possible combinations of common and distinct components based on a total rank of three. The horizontal axis represents the number of common components and the numbers in the plot represent the number of distinct components for SMELL and TASTE respectively. c) PCA-GCA: black dots represent the canonical correlation coefficient (x100) and the bars show how much variance the canonical components explain in each block.}
 \label{Fig_SensorExplVar}
\end{figure}

\noindent A crucial aspect of the decomposition is to decide the dimensions of the common and distinct subspaces. For DISCO, this is a two-step process: first, the number of SCA components is selected. This number represents the sum of the dimensions of all subspaces, i.e. $dim R(\bX_{12C})+ dim R(\bX_{1D})+ dim R(\bX_{2D})$. Then, the most appropriate target matrix $\bP_{Target}$ (Eqn. \ref{eDISCO2}) is sought by evaluating the non-congruence value (Eqn. \ref{eDISCO1}) for all possible allocations of common and distinct components. Since there are three independent design factors in this experiment (flavor type, flavor dose and sugar level), we choose to keep three SCA components even if the third component explain very little variance (Figure \ref{Fig_SensorExplVar}a). The lowest non-congruence value is approximately equal for models with one and two common components (Figure \ref{Fig_SensorExplVar}b), but after a closer inspection of the scores we choose the model with one common component and one disticnt component per block.\\

\noindent For real data the non-congruence value is never zero, meaning that the zeros in the target matrix are not exactly zero in the rotated loadings, which means that the distinct component for one block also explains some variance in the other block. The latter is the distinct-non-orthogonal subspace. The DISCO decomposition for this data set is then:
\begin{eqnarray}
\label{eSENSORY1}
    \bX_{S} &=& \bX_{C,S} (70.0\%) + \bX_{D,T} (3.4\%) + \bX_{D,S} (9.4\%) + \bE_S \\ \nonumber
    \bX_{T} &=& \bX_{C,T} (26.8\%) + \bX_{D,T} (55.5\%) + \bX_{D,S} (1.9\%) + \bE_T
\end{eqnarray}
where each subspace is of dimension one; S and T stand for Smell and Taste, respectively and between brackets is the amount of explained variation in the block. Note that both distinct-non-orthogonal subspaces are very small in this case, and probably consist of noise only.\\

\noindent For PCA-GCA, the dimension selection is also a stepwise procedure: First, an appropriate number of principal components is selected for each data block, corresponding to $dim R(\bX_1)$ and $dim R(\bX_2)$ in Eqn \ref{eDim1}. Next, the correlation coefficients and explained variances from GCA are evaluated in order to decide the number of common components, $dim R(\bX_{12C})$. The number of distinct components is then given as the difference between $dim R(\bX_k)$ and $dim R(\bX_{12C})$. In this example, we choose to keep three components for each block, following the same argument as for DISCO (three design factors). Figure \ref{Fig_SensorExplVar}c shows that the canonical correlation together with the explained variances clearly suggest one common component (correlation = 0.98), which means that the distinct subspaces are two-dimensional. The distinct subspaces can be split into an orthogonal and non-orthogonal part as for DISCO, but that is not done here. The decomposition from PCA-GCA is then:
\begin{eqnarray}
\label{eSENSORY2}
    \bX_{S} &=& \bX_{C,S} (86.4\%) + \bX_{D,S} (9.3\%) + \bE_S \\ \nonumber
    \bX_{T} &=& \bX_{C,T} (25.7\%) + \bX_{D,T} (66.2\%) + \bE_T
\end{eqnarray}
where the common part has dimensionality one and both distinct parts have dimensionality two.\\

\begin{figure}[h!]
 \centerline{\includegraphics*[width=14cm]{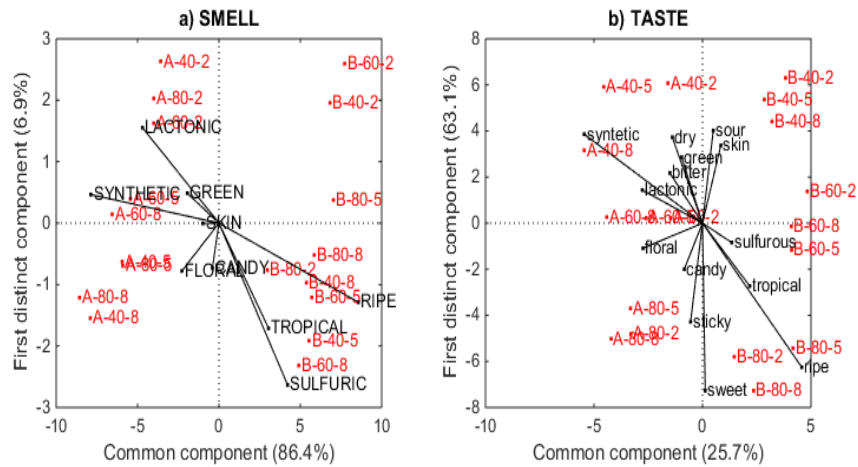}}
 \caption{\footnotesize : Biplots from PCA-GCA, showing the variables as vectors and the samples as
points. The samples are labeled according to the design factors flavor type (A/B), sugar level
(40,60,80) and flavor dose (2,5,8). The plots show the common component (horizontal) against
the first distinct component for each of the two blocks.}
 \label{Fig_SensorBiplot}
\end{figure}

\noindent The subspaces found by PCA-GCA and DISCO are very similar. The correlation between the common DISCO component ($\bT_{12C}$) and the common PCA-GCA components ($\bT_{1C}$ and $\bT_{2C}$) are 0.98 for both blocks. The correlation between the distinct (orthogonal) SMELL component from DISCO ($\bT_{1DO}$) and the first distinct SMELL component from PCA-GCA (first column of $\bT_{2D}$) is 0.74. The corresponding number for the distinct TASTE components is 0.99. Figure \ref{Fig_SensorBiplot} shows biplots from PCA-GCA for each of the two blocks. It is clear that the common component distinguishes between flavor type (A and B). This component explains 86\% of the SMELL variation and 26\% of the TASTE variation. As a validation of the commonness, note that the sensory attributes that span this subspace are the same both for smelling and tasting: synthetic/lactonic/oral for flavor A versus ripe/tropical/sulfurous for flavor type B. The first distinctive SMELL component explains 7\% of the variation and is related to the flavor dose, showing that the lowest dose tend to give a more lactonic smell. The first distinctive TASTE component explains 63\% of the variation and describes differences in sugar level. The attributes that span this component are sweet/ripe versus sour/synthetic/skin/dry.\\

\noindent This example shows that both methods are able to separate common and distinct subspaces in a similar way. The subspaces that explain a large proportion of the variance (common and distinct TASTE) are practically equal for both methods (correlations $>0.98$), while there is less agreement regarding the weaker distinct SMELL component (correlation = 0.74).

\subsection{Medical biology example}

\noindent The data set is a subset of a larger study on the effects of gastric bypass surgery on obese and diabetic subjects  \cite{Lips2014}. Here, we focus on 14 obese patients with Diabetes Mellitus Type II (DM2) who underwent gastric bypass surgery. Blood samples were taken four weeks before and three weeks after surgery and on each occasion samples were taken both before and after a meal. The blood samples were then analyzed on multiple analytical platforms for the determination of amines, lipids and oxylipins. The three data blocks Amines (A), Lipids (L) and Oxylipins (O) consist of 14 subjects x 4 samples = 56 rows, and 34, 243 and 32 variables respectively. All variables in all three blocks were square-root transformed, in order to obtain more evenly distributed data. Individual differences between subjects were removed by subtracting each subjects' average profile. All variables were then scaled to unit variance. The blocks were also scaled to unit norm prior to SCA, to normalize scale differences between blocks.\\

\begin{figure}[h!]
 \centerline{\includegraphics*[width=16cm]{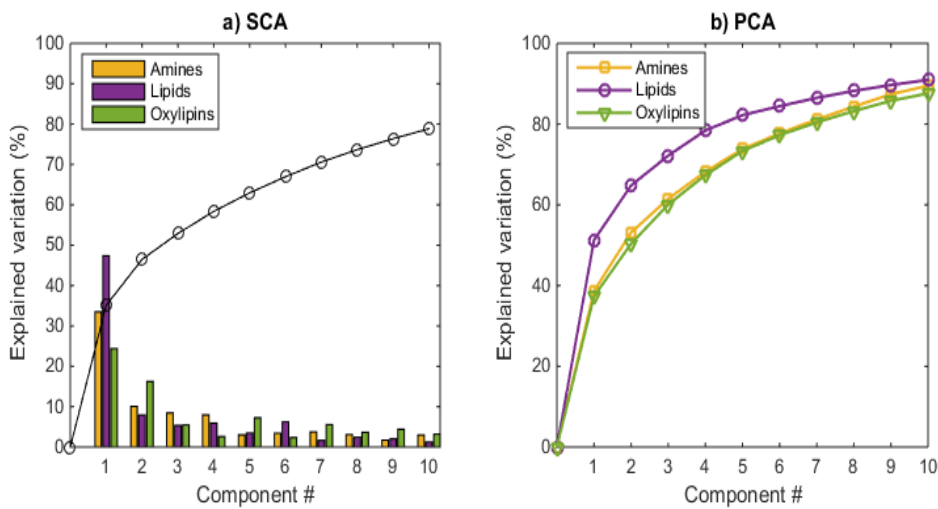}}
 \caption{\footnotesize Explained variances for a) SCA. The bars represent variances within each block, and the curve represent cumulative explained variance in all blocks combined. b) PCA on each block separately.}
 \label{Fig_MetExplVar}
\end{figure}

\begin{figure}[h!]
 \centerline{\includegraphics*[width=16cm]{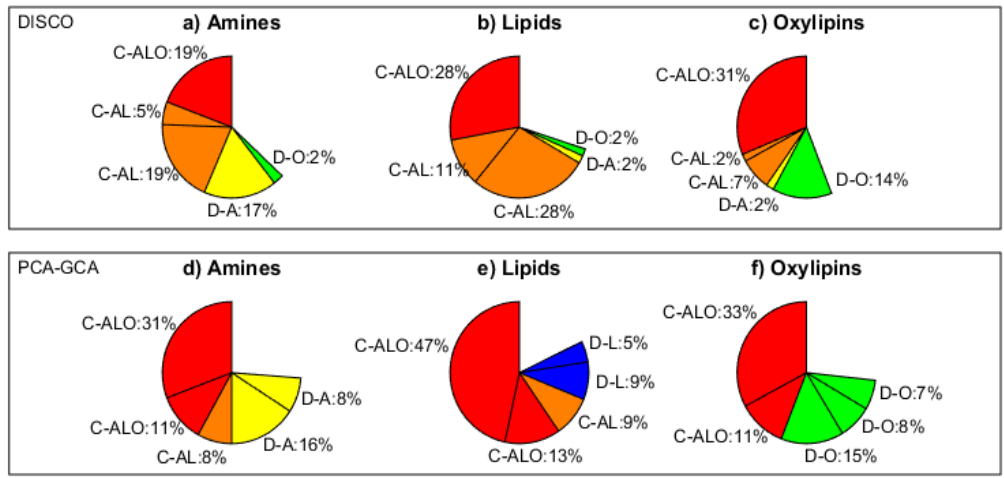}}
 \caption{\footnotesize Subplots a), b) and c) show the decomposition by DISCO for blocks A, L and O respectively, while d), e) and f) show the corresponding decomposition by PCA-GCA. Each segment represent a component (dimension).}
 \label{Fig_DISCOExplVar}
\end{figure}

\noindent Selecting the dimensions of the subspaces is more complicated when the numbers of blocks increase. In this three-block example, we need to decide the dimensions of seven subspaces: $\bX_{123C}$, $\bX_{12C}$, $\bX_{13C}$, $\bX_{23C}$, $\bX_{1D}$, $\bX_{2D}$, and $\bX_{3D}$. For DISCO, we start by deciding the sum of all the dimension, i.e. the number of SCA components. Explained variance as a function of components for SCA is given in Figure \ref{Fig_MetExplVar}a. The curve of cumulative variance does not have a clear bend, which makes it hard to decide the cutoff between structure and noise. To allocate the common and distinct components we need to fix the number of SCA components and then compare the fit values of Eqn. \ref{eDISCO1}. The computations are time-consuming, as there are e.g. 462 possible target matrices for the 5-component model. To illustrate the complexity in selecting the dimensions for the subspaces, we have calculated all possible rotations for models with 3-5 SCA components, and the results for the four best-fit values are given in Table \ref{Table3}. The values are very similar, making it hard to conclude which rotation gives the best fit. Looking further into the actual rotated score vectors, we discover that many of the models agree on some of the subspaces. These are marked with colors in Table \ref{Table3}. We choose to interpret the 5-component model with fit value 0.24 (the best 5-component model), since this model includes all the agreed upon subspaces. The model contains one component that is common across all three blocks, two components common for A and L, and one distinct component from both A and O. The decomposition of each block is illustrated by pie charts in Figure \ref{Fig_DISCOExplVar}a-c. Notice that there is a substantial contribution of one of the C-AL components also in the O block (7\%), which implies that this component could perhaps also be regarded as common across all three blocks.\\

\noindent In PCA-GCA, the number of principal components need to be set for each block separately before performing GCA. Explained variations for the three PCA models are shown in Figure \ref{Fig_MetExplVar}b. As for SCA, it is not clear how many components to keep for each block. To investigate how the choice affect the GCA, we ran GCA on all combinations of 5-8 components from each block (64 combinations in total). The canonical correlation coefficient for cases with more than two blocks is defined as the average correlation between all pairs of components from different blocks. Using 0.7 as correlation threshold for commonness in the GCA, we found that 85\% of the models had two common components across all blocks, and one common component across A and L. The model based on five components for each block is illustrated in Figure \ref{Fig_DISCOExplVar}d-f. Closer investigation of the components revealed that the second common component across all three blocks is very similar to the one of the C-AL DISCO component mentioned above, which explained 7\% of the variation in O. This illustrates the complexity of splitting common and distinct components in noisy and complex data.\\

\begin{figure}[h!]
 \centerline{\includegraphics*[width=16cm]{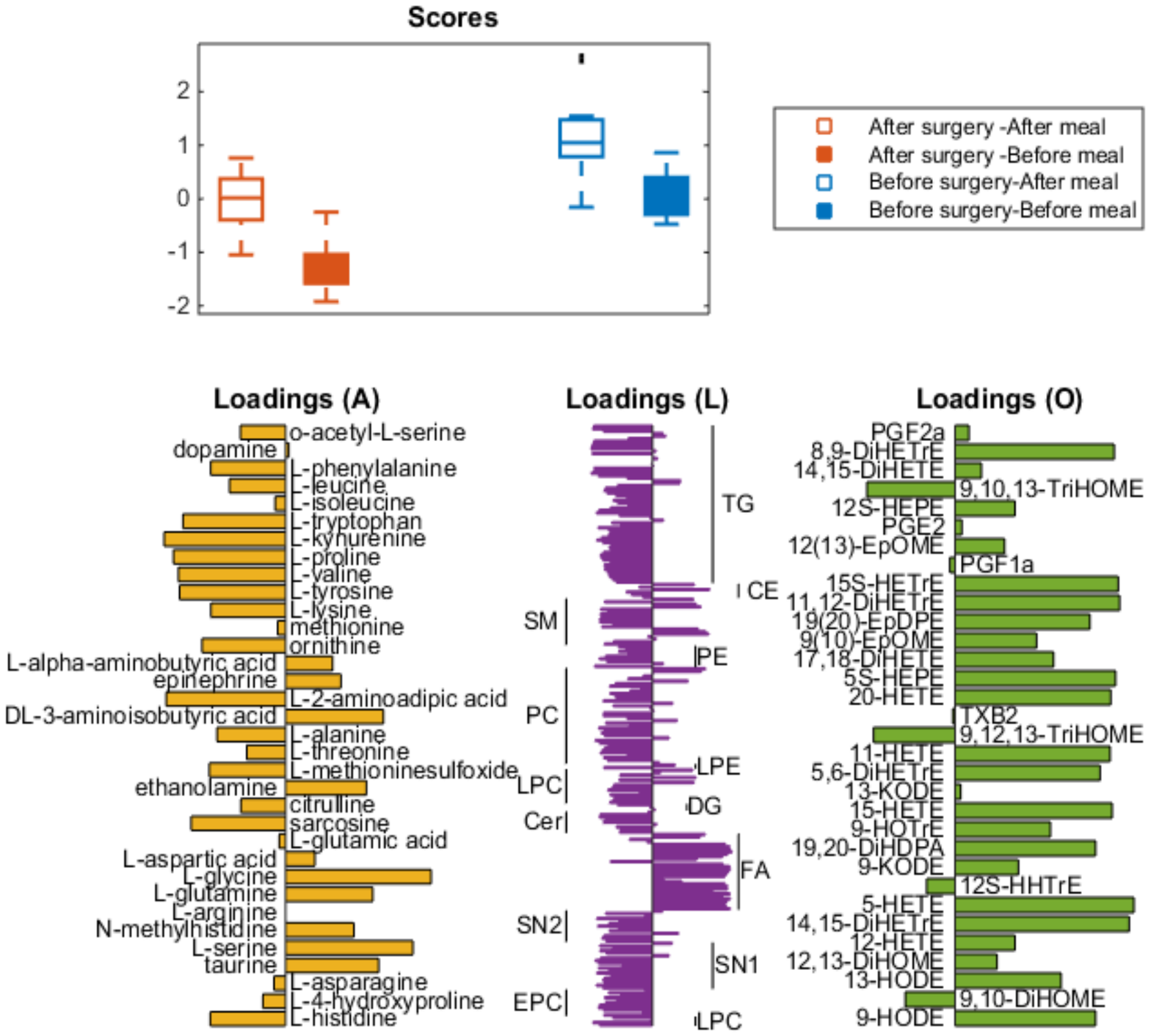}}
 \caption{\footnotesize Scores and loadings for the one-dimensional DISCO subspace common between all three blocks.}
 \label{Fig_MetALO}
\end{figure}

\noindent To interpret the different subspaces, we plot the scores and loadings from the DISCO model. Figure \ref{Fig_DISCOExplVar} shows the one-dimensional subspace that is common for all three blocks (C-ALO), which accounts for 19\%, 28\% and 31\% of the variation in A, L and O respectively. The scores are shown in the top panel of Figure \ref{Fig_DISCOExplVar}. It is clear that the component contains information both related to surgery and meal; the scores are increasing after surgery and decreasing after the meal. The variables spanning this dimension in each of the three blocks are shown in the bar plots of Figure \ref{Fig_DISCOExplVar} (bottom). The most striking observation is that the branched chain amino acids leucine, valine (and to a lesser extend leucine) and L-2-aminoadipic acid (closely related to branched chain amino acids) are down regulated after surgery, which confirms earlier findings \cite{Lips2014}. There is more in common between amines and lipids than oxylipids; both amines and lipids are involved in central carbon and energy metabolism and therefore they may show higher correlation among some amino acids and some lipid groups (as reflected by common subspace).\\

\begin{figure}[h!]
 \centerline{\includegraphics*[width=16cm]{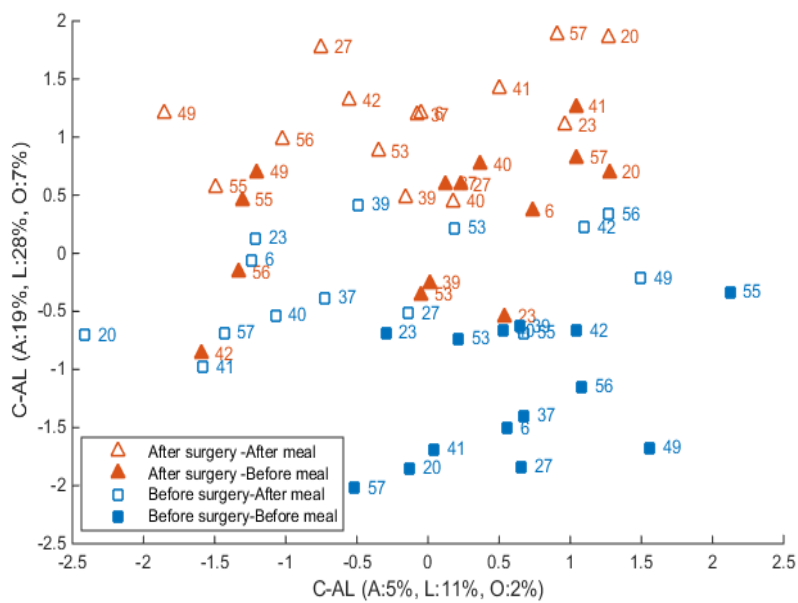}}
 \caption{\footnotesize Scores for the two-dimensional DISCO subspace common between the amine and lipids blocks.}
 \label{Fig_MetAL12}
\end{figure}

\noindent The two-dimensional subspace common between A and L is shown in Figure \ref{Fig_MetAL12}. These two components together account for 24\% and 39\% in the A and L blocks respectively, and they even explain 9\% in the O block. Here also, we see groupings according to both surgery and meal, especially in the vertical dimension. Note that the two groups that were overlapping in the C-ALO component ("before surgery-before meal" versus "after surgery-after meal") are completely separated in this subspace. Plots of the distinct components (not shown) did not reveal clear patterns related to the factors treatment and meal. Hence, all effects are seen in the common parts meaning that a large part of the metabolism is affected simultaneously by these two factors.

\section{Discussion}

\subsection{Revisiting the framework}

After having given a short tour of methods for finding common and distinct components, it is worthwhile to recapitulate the general mathematical framework as presented in Eqns. \ref{eDirectSum2} and \ref{eDirectSum4}. A summary of the models underlying the presented methods is given in the column 'Model' of Table \ref{Table2}. It appears that for the two-block case the general model is
\begin{eqnarray}\label{eGenModel}
    \bX_1 &=& \bX_{1C} + \bX_{1D} + \bE_1 \\ \nonumber
    \bX_2 &=& \bX_{2C} + \bX_{2D} + \bE_2
\end{eqnarray}
with different properties of the matrices $\bX_{kC}$, $\bX_{kD}$ and $\bE_k$. Some methods do not estimate $\bX_{kD}$ (GCA) and some methods do not distinguish between common and distinct (SCA). Different choices are made regarding the positioning of the column-spaces of $\bX_{kC}$ and $\bX_{kD}$ (see Table \ref{Table2} under 'Subspace properties'). Also different (although sometimes implicit) choices are made regarding orthogonality (see Table \ref{Table2} under 'Orthogonality') resulting in differences in (E)SS. None of the methods makes a rigorous direct sum decomposition as in Eqns. \ref{eDirectSum2} and \ref{eDirectSum4}. The cases for more than two blocks shows an even wider variety of possibilities. The choice of orthogonality constraints depends on the application. It may well be that in most practical applications only the common and orthogonal distinct parts are the most informative.

\subsection{Finding common and distinct subspaces}

There is an interesting difference in the way the various methods find common and distinct subspaces. Some methods work clearly in the column-spaces of the matrices involved (GCA, JIVE), some methods work through the row-spaces (DISCO, O2PLS) and some methods work in both types of spaces simultaneously (GSVD and Structure Revealing Data Fusion). Whether or not this has consequences for the interpretation of the results of the different models is an open question.

\subsection{Open issues and future work}

There are obviously many open issues in this field of research. We have only briefly touched upon the issue of explained variances, but there are many nontrivial aspects that need attention. Also the problem of interpretation, that is, moving from layer three to four in Figure \ref{Fig_Cartoon}, needs attention. This is a very important issue because interpretation is one of the raisons-d'\^{e}tre for data fusion methods. Our framework primarily considers the column space of the data matrices but interpretation is done mostly in the row-space. How to investigate this depends on the scope of the analysis and the type of data available and it is not possible to set up a completely general procedure. There are, however, some general tools that can be useful:  One important possibility is to simply project the original data blocks onto the estimated subspaces. For instance, for $R(\bX_{12C})$ one simply regresses $\bX_1$ onto a suitable basis for the space $R(\bX_{12C})$. In this way one obtains information about how the original data are related to the basis for each subspace. Also moving to analyzing more than two blocks simultaneously is not trivial. Many choices have to be made and no clear guidelines exist on how to perform this. Model selection becomes an even more important issue then and possibly Bayesian factor analysis methods with automated model selection can be of use in this context \cite{Ray2014}.

\section{Acknowledgements}

We thank Frans van der Kloet and Johan Westerhuis (both from Biosystems Data Analysis, University of Amsterdam) for stimulating discussions.

\section{Appendix}

\subsection{Possibilities of orthogonal decompositions}
\label{Possibilities of orthogonal decompositions}

\subsubsection{Choices for Distinct-Orthogonal (DO) spaces.}

There are several possibilities for choosing orthogonality in the decompositions of Eqn. \ref{eDirectSum2}. These will be outlined and explained below.
We will focus attention on $R(X_{1DO})$ but analogous results hold for $R(X_{2DO})$; we will consider $R(X_{1DNO})$ as a 'rest' term and not consider this space explicitly. The first level to discuss possibilities is regarding the status of $R(X_{1D})$. The alternatives are:
\begin{description}
  \item[A0:] no orthogonality restrictions for $R(X_{1D})$.
  \item[A1:] $R(X_{1D}) \bot R(X_{12C})$ (see Figure \ref{Fig_Distinct}a).
  \item[A2:] $R(X_{1D}) \bot R(X_{2D})$ (see Figure \ref{Fig_Distinct}b).
  \item[A3:] $R(X_{1D}) \bot R(X_2)$ which implies \textbf{A1} and \textbf{A2}.
\end{description}
and, as said earlier, alternative \textbf{A3} is not always possible. The status of $R(X_{1DO})$ is nested in alternatives \textbf{A0-A3}, since $R(X_{1DO})$ is a part of $R(X_{1D})$.\\

\noindent The alternatives under \textbf{A0} are:
\begin{description}
  \item[A01:] $R(X_{1DO}) \bot R(X_{2DO})$.
  \item[A02:] $R(X_{1DO}) \bot R(X_{2D})$.
  \item[A03:] $R(X_{1DO}) \bot R(X_2)$.
\end{description}
which shows increasing degrees of orthogonality.\\

\noindent The alternatives under \textbf{A1} are:
\begin{description}
  \item[A10:] $R(X_{1DO}) \bot R(X_{12C})$ which follows from \textbf{A1}.
  \item[A11:] $R(X_{1DO}) \bot R(X_{12C})$ and $R(X_{1DO}) \bot R(X_{2DO})$.
  \item[A12:] $R(X_{1DO}) \bot R(X_{12C})$ and $R(X_{1DO}) \bot R(X_{2D})$ which implies $R(X_{1DO}) \bot R(X_2)$ and is the same as alternative \textbf{A03}.
\end{description}
which shows again increasing degrees of orthogonality.\\

\noindent The alternatives under \textbf{A2} are:
\begin{description}
  \item[A20:] $R(X_{1DO}) \bot R(X_{2D})$ which follows from \textbf{A2} and is the same as alternative \textbf{A02}.
  \item[A21:] $R(X_{1DO}) \bot R(X_{2D})$ and $R(X_{1DO}) \bot R(X_{12C})$ which is again the same as alternative \textbf{A03}.
\end{description}
and under alternative \textbf{A3} there is only one option namely $R(X_{1DO}) \bot R(X_2)$ which is again the same as option \textbf{A03}. Concluding, for the two block case there are five different alternatives to select $R(X_{1DO})$: \textbf{A01}, \textbf{A02}, \textbf{A03}, \textbf{A10} or \textbf{A11}. Whether these alternatives are available for a specific application depends on the dimensions and positioning of the subspaces. An example of this and how to analyze such situations is presented in the next Subsection \ref{A specific example}.\\

\subsubsection{A specific example.}
\label{A specific example}

As an example of using rigorous linear algebra results to explore possibilities for (non-)orthogonal decompositions consider the example of two distinct subspaces both of dimension two (see Section \ref{The two-block case}). It can be proven that if $R(\bX_{1D})$ and $R(\bX_{2D})$ are both two-dimensional and not orthogonal, then for every vector $\bx$ in $R(\bX_{1D})$ there is exactly one vector $\by$ in $R(\bX_{2D})$ orthogonal to $\bx$. This goes as follows. Suppose that $\bA$ and $\bB$ are both orthogonal matrices serving as bases for $R(\bX_{1D})$ and $R(\bX_{2D})$, respectively. Assume also that $r(\bA^T\bB)=2$ where $r(.)$ means the rank of a matrix. This implies:
\begin{itemize}
  \item $R(\bX_{1D})$ is not orthogonal to $R(\bX_{2D})$; otherwise $\bA^T \bB = 0$
  \item $R(\bX_{1D})$ does not contain a vector orthogonal to the whole of $R(\bX_{2D})$ (this vector $\bg$ could be written as $\bg = \bA \bh$ and then $\bh^T \bA^T \bB =0$, which contradicts $r(\bA^T \bB)=2$)
  \item $R(\bX_{2D})$ does not contain a vector orthogonal to the whole of $R(\bX_{1D})$ (analogously as above)
\end{itemize}
now there is for any nonzero vector $\bx \in R(\bX_{1D})$ exactly one nonzero vector $\by \in R(\bX_{2D})$ such that $\bx^T\by=0$.\\

\noindent Proof: write $\bx = \bA \bu$, $\by = \bB \bv$ and $\bB = \bA \bU + \bA^\perp \bV$. Find a vector $\bv$ such that $\bx^T\by=0$, or, such that $\bu^T \bA^T(\bA \bU + \bA^\perp \bV)\bv=0$ which equals $\bu^T \bU\bv=0$ because of the orthogonality of $\bA$ and the definition of the orthogonal complement $\bA^\perp$. Then $\bU=\bA^T \bB$ which follows by pre-multiplying $\bB = \bA \bU + \bA^\perp \bV$ with $\bA^T$; by defining $\bt = [t_1|t_2]^T = \bU^T \bu$ which is a unique nonzero vector (because $r(\bU=\bA^T \bB)=2$) it follows that $\bv = [t_2|-t_1]^T$ is the vector which makes $\bu^T \bU\bv=0$. Hence, there is exactly one vector $\by=\bB \bv$ in $R(\bX_{2D})$ which is orthogonal to $\bx$.

\subsection{GCA proof}

In the main text it was stated that the solution $\bT$ in Eqn. \ref{eGCA1} is in the column space of $[\bX_1|...|\bX_K]$. This will be proven now for the two-block situation for simplicity, but is easily generalized to the more-than-two block situation. Eqn. \ref{eGCA2} can also be written as
\begin{equation}\label{eGCA3}
    \sum^2_{k=1} \bX_k (\bX^T_k \bX_k)^+ \bX^T_k = [\bX_1|\bX_2] [\bX_1(\bX^T_1\bX_1)^+|\bX_2(\bX^T_2\bX_2)^+]^T = \bT \bS \bT^T
\end{equation}	
where the full eigenvalue decomposition (i.e. $\bS>0$) is used. Post-multiplying both sides of Eqn. \ref{eGCA3} by $\bT \bS^{-1}$ gives now
\begin{equation}\label{eGCA4}
    [\bX_1|\bX_2] [\bX_1(\bX^T_1\bX_1)^+|\bX_2(\bX^T_2\bX_2)^+]^T \bT \bS^{-1} = \bT
\end{equation}	
or
\begin{equation}\label{eGCA5}
    [\bX_1|\bX_2] \bQ = \bT
\end{equation}	
which shows that $\bT$ (and also its first columns when only those are used) is in the range of $ [\bX_1|\bX_2]$. This argument is easily extended to more than two blocks.

\subsection{Relationship between O2PLS and Procrustes Analysis}

An interesting relationship exists between O2PLS and Procrustes Analysis. The Procrustes problem can be stated as
\begin{equation}\label{eProc1}
    \min_{\bR^T \bR = \bI} \|\bX_2 \bR -\bX_1\|^2
\end{equation}	
and the solution of this problem is $\bR = \bU \bV^T$ where $\bU$ and $\bV$ are from the SVD of $\bX^T_2 \bX_1 = \bU \bS \bV^T$ \cite{Schonemann1966}. Then post-multiplying both $\bX_2 \bR$ and $\bX_1$ with $\bV$ gives $\bX_2 \bR \bV = \bX_2 \bU \bV^T \bV = \bX_2 \bU$ and $\bX_1 \bV$ which are the same quantities as obtained for O2PLS. Note that the Procrustes problem of Eqn. \ref{eProc1} is equivalent to
\begin{equation}\label{eProc2}
    \min_{\bR_k^T \bR_k = \bI} \|\bX_2 \bR_2 -\bX_1 \bR_1\|^2
\end{equation}	
which is the symmetric formulation of the problem with solution $\bR_2=\bU$ and $\bR_1=\bV$ \cite{TenBerge1977}.

\section{Tables}

\renewcommand{\arraystretch}{2}

\small
\begin{landscape}
\begin{longtable}[c]{@{}p{4cm}p{12cm}p{4cm}@{}}
\caption{\textbf{Fundamental aspects of the fusion methods}\\ Abbreviations: C is common; D is distinct. Colors: brown is mixed subspace; green is common subspace; red/blue are distinct subspaces.} \label{Table1}  \\
\mytoprule
Methods & Model & Uniqueness \\
\midrule
SCA & $\bX_k=\textcolor{brown}{\bT \bP^T_k}+\bE_k$ or \newline $\bX_k=\textcolor{brown}{\widetilde{\bT}_k \bP^T_k}+\bF_k$ ($\widetilde{\bT}_k=\bX_k\bX^+_k\bT $)& $R(\bT)$ is unique \\ \hdashline
GCA & $\bX_k=\textcolor{green}{\bX_k \bW_k \bP^T_k}+\bE_k=\textcolor{green}{\bT_k \bP^T_k}+\bE_k=\textcolor{green}{\bX_{kC}}+\bE_k$ & $R(\bT)$ is unique \\ \hdashline
O2PLS & $\bX_1=\textcolor{green}{\bT_{1C} \bW^T_1}+\textcolor{red}{\bT_{1D}\bP_{1D}^T}+\bE_1=\textcolor{green}{\bX_{1C}}+\textcolor{red}{\bX_{1D}}+\bE_1$ \newline $\bX_2=\textcolor{green}{\bT_{2C} \bW^T_2}+\textcolor{blue}{\bT_{2D}\bP_{2D}^T}+\bE_2=\textcolor{green}{\bX_{2C}}+\textcolor{blue}{\bX_{2D}}+\bE_2$ & ?? \\ \hdashline
DISCO & $\bX_1=\textcolor{green}{\bT_1\bP_{11}^T}+\textcolor{red}{\bT_2\bP_{12}^T}+\textcolor{red}{\bT_3\bP_{13}^T}+\bE_1=\textcolor{green}{\bX_{1C}}+\textcolor{red}{\bX_{1DO}}+\textcolor{red}{\bX_{1DNO}}+\bE_1$ \newline
$\bX_2=\textcolor{green}{\bT_1\bP_{21}^T}+\textcolor{blue}{\bT_2\bP_{22}^T}+\textcolor{blue}{\bT_3\bP_{23}^T}+\bE_2=\textcolor{green}{\bX_{2C}}+\textcolor{blue}{\bX_{2DNO}}+\textcolor{blue}{\bX_{2DO}}+\bE_2$ & $R(\bT)$ is unique \\ \hdashline
GSVD & $\bX_1=\bT \bD_1 \bV^T_1+\bE_1=\textcolor{red}{\bT_1 \bD_{11} \bV^T_{11}}+\textcolor{green}{\bT_2 \bD_{12} \bV^T_{12}}+\textcolor{red}{\bT_3 \bD_{13} \bV^T_{13}}+\bE_1$ = \newline $\textcolor{red}{\bX_{1DO}}+\textcolor{green}{\bX_{1C}}+\textcolor{red}{\bX_{1DNO}}+\bE_1$ \newline $\bX_2=\bT \bD_2 \bV^T_2+\bE_2=\textcolor{blue}{\bT_1 \bD_{21} \bV^T_{21}}+\textcolor{green}{\bT_2 \bD_{22} \bV^T_{22}}+\textcolor{blue}{\bT_3 \bD_{23} \bV^T_{23}}+\bE_2$ = \newline  $\textcolor{blue}{\bX_{2DNO}}+\textcolor{green}{\bX_{2C}}+\textcolor{blue}{\bX_{2DO}}+\bE_2$ & See Section \ref{GSVD} \\ \hdashline
JIVE & $\bX_k=\textcolor{green}{\bT (\bP_{kC}^T})+\textcolor{red}{\bT_k (\bP_{kD})^T}+\bE_k=\textcolor{green}{\bX_{kC}}+\textcolor{red}{\bX_{kD}}+\bE_k$ & Subspaces unique \\ \hdashline
SRDF & $\bX_k=\bT \bD_k \bV^T_k+\bE_k=\textcolor{green}{\bX_{kC}}+\textcolor{red}{\bX_{kD}}+\bE_k$ & ?? \\ \hdashline
\mybottomrule
\end{longtable}
\end{landscape}

\thispagestyle{empty}

\begin{landscape}
\begin{longtable}[c]{@{}p{1.5cm}p{9.5cm}p{7cm}@{}}
\caption{\textbf{Some properties of the fusion methods}} \label{Table2}  \\
\mytoprule
Methods & Subspace properties & Orthogonality \\
\midrule
SCA                                                                                                    &
$R(\bT)\subseteq R[\bX_1|...|\bX_K]$;  $R(\bT) \nsubseteq R[\bX_k]$ ;\, $R(\widetilde{\bT}_k)\subseteq R(\bX_k)$   &
$\bT^T\bT=\bI$                                                                                                                    \\
GCA                                                                                                    &
$R(\bT)\subseteq R[\bX_1|...|\bX_K]$; $R(\bT)\nsubseteq R[\bX_k]$; $R(\widetilde{\bT}_k)\subseteq R(\bX_k)$          &
$\bT^T\bT=\bI $                                                                                                                    \\
O2PLS                                                                                                  &
$R(\bX_{kC}) \subseteq R(\bX_k)$; $R(\bX_{kD}) \subseteq R(\bX_k)$                                     &
$\bX_{kC}^T \bX_{kD}=0$; $\bE_k^T \bX_{kD}=0$; $\bE_k^T \bX_{kC} \neq 0$; $\bX_{kC}^T \bX_{k'D} \neq 0 (k \neq  k')$; $\bX_{kD}^T \bX_{k'D} \neq 0 (k \neq  k')$ \\
DISCO                                                                                                  &
$R(\bT)\subseteq R[\bX_1|...|\bX_K]$ $R(\bT_l) \nsubseteq R(\bX_k);l=1,2,3;\forall k$                              &
All orthogonal except:
$\begin{array}[t]{rll@{}} (\bX_{1DO})^T \bX_{2DNO} &\neq & 0; \\
                          (\bX_{2DO})^T \bX_{1DNO} &\neq & 0
\end{array}$                                                                                       \\
GSVD                                                                                                   &
$R(\bT)\subseteq R[\bX_1|\bX_2]$; $R(\bT_l)\nsubseteq R(\bX_k);\,\,l=1,2,3;k=1,2$                             &
$\bT^T \bT\neq0$; $\bX_{1DO}$, $\bX_{1DNO}$ and $\bX_{1C}$ mutually orthogonal; $\bX_{2DO}$, $\bX_{2DNO}$ and $\bX_{2C}$ mutually orthogonal \\
JIVE                                                                                                   &
$R(\bT)\nsubseteq R(\bX_k)$ ; $R(\bT_k)\subseteq R(\bX_k)$ ; $R(\bT)\subseteq R[\bX_1|...|\bX_K]$                                           &
$(\bX_{k'C})^T \bX_{kD}=0;\forall k,k'$; $(\bX_{k'D})^T \bX_{kD} \neq 0 (k \neq  k')$                  \\
SRDF                                                                                                   &
??                                          &
No orthogonality                                                                                          \\
\mybottomrule
\end{longtable}
\end{landscape}

\normalfont
\newpage

\renewcommand{\arraystretch}{1}

\begin{table}[h!]
\caption{Overview of DISCO models for the medical biology example. The table shows models for the four lowest fit-values for rotations based on 3-5 SCA-components. Components are labeled as common (C-) or distinct (D-). The colored components are subspaces that are the same across several models and the correlation between these subspaces are given in Table 3. The framed model is selected for further interpretation.} \label{Table3}
\begin{tabular}[t]{llllllllll}
\toprule
 & \multicolumn{8}{l}{Increasing Fit Values} & \\
\midrule
SCA comp & \multicolumn{2}{l}{1} & \multicolumn{2}{l}{2} & \multicolumn{2}{l}{3} & \multicolumn{2}{l}{4} & ExplVar \\
\midrule
\multirow{3}*{3} & \textcolor{green}{\textbf{C-AL}} & \multirow{3}*{0.13} & C-AL & \multirow{3}*{0.15} & C-AO & \multirow{3}*{0.16} & C-ALO & \multirow{3}*{0.20} & \multirow{3}*{53\%} \\
 & \textcolor{green}{\textbf{C-AL}} &  & C-AO &  & C-ALO &  & C-ALO &  &  \\
 & \textcolor{violet}{\textbf{C-ALO}} & & C-LO & & \textcolor{violet}{\textbf{C-ALO}} & & C-ALO & & \\
\midrule
\multirow{4}*{4} & C-ALO & \multirow{4}*{0.19} & \textcolor{red}{\textbf{D-A}} & \multirow{4}*{0.20} & \textcolor{red}{\textbf{D-A}} & \multirow{4}*{0.23} & \textcolor{green}{\textbf{C-AL}} & \multirow{4}*{0.24} & \multirow{4}*{58\%} \\
 & C-ALO & & \textcolor{green}{\textbf{C-AL}} & & D-O & & \textcolor{green}{\textbf{C-AL}} & & \\
 & C-ALO & & \textcolor{green}{\textbf{C-AL}} & & C-ALO & & C-AL & & \\
 & C-ALO & & \textcolor{violet}{\textbf{C-ALO}} & & C-ALO & & \textcolor{violet}{\textbf{C-ALO}} & & \\
\midrule
\multirow{5}*{5} & \textcolor{red}{\textbf{D-A}} & \multirow{5}*{0.24} & C-AL & \multirow{5}*{0.26} & \textcolor{red}{\textbf{D-A}} & \multirow{5}*{0.28} & \textcolor{red}{\textbf{D-A}} & \multirow{5}*{0.28} & \multirow{5}*{63\%} \\
 & \textcolor{blue}{\textbf{D-O}} & & C-AO & & \textcolor{blue}{\textbf{D-O}} & & \textcolor{blue}{\textbf{D-O}} & & \\
 & \textcolor{green}{\textbf{C-AL}} & & C-ALO & & C-AL & & C-LO & & \\
 & \textcolor{green}{\textbf{C-AL}} & & C-ALO & & C-ALO & & C-ALO & & \\
 & \textcolor{violet}{\textbf{C-ALO}} & & \textcolor{violet}{\textbf{C-ALO}} & & \textcolor{violet}{\textbf{C-ALO}} & & \textcolor{violet}{\textbf{C-ALO}} & & \\
\bottomrule
\end{tabular}
\end{table}

\bibliographystyle{plain}
\bibliography{CommonDistinct_v1}

\end{document}